\documentclass[12pt]{article}
\usepackage{graphicx}
\usepackage{pslatex}
\usepackage[latin1]{inputenc}
\usepackage[T1]{fontenc}
\usepackage{txfonts}
\usepackage{latexsym}
\usepackage{cite}
\usepackage{pstricks,rotating, graphicx}

\usepackage{color}
\usepackage{colordvi}

\graphicspath{{figs/}}
\DeclareGraphicsExtensions{.eps,.ps}

\makeatletter
\def\@sect#1#2#3#4#5#6[#7]#8{\ifnum #2>\c@secnumdepth
  \def\@svsec{}\else 
  \refstepcounter{#1}\edef\@svsec{\csname the#1\endcsname \hskip0.5em}\fi
  \@tempskipa #5\relax
  \ifdim \@tempskipa>\z@
    \begingroup 
      #6\relax
      \@hangfrom{\hskip #3\relax\@svsec}{\interlinepenalty \@M #8\par}%
    \endgroup
    \csname #1mark\endcsname{#7}\addcontentsline
      {toc}{#1}{\ifnum #2>\c@secnumdepth \else
        \protect\numberline{\csname the#1\endcsname}\fi #7}%
  \else
    \def\@svsechd{#6\hskip #3\@svsec #8\csname #1mark\endcsname
      {#7}\addcontentsline{toc}{#1}{\ifnum #2>\c@secnumdepth \else
        \protect\numberline{\csname the#1\endcsname}\fi #7}}%
  \fi \@xsect{#5}}
\@addtoreset{equation}{section}
\makeatother

\renewcommand\theequation{\ifnum \value{section}>0
 \arabic{section}.\arabic{equation}%
\else
\arabic{equation}%
\fi}

\def\Eq#1{Eq.~(\ref{#1})}
\def\Eqs#1#2{Eqs.~(\ref{#1})--(\ref{#2})}
\def\NLO{{\mbox{\scriptsize NLO}}}
\def\LO{{\mbox{\scriptsize LO}}}
\def\virt{{\mbox{\scriptsize virt}}}
\def\real{{\mbox{\scriptsize real}}}
\def\fac{{\mbox{\scriptsize fact}}}
\def\Dipole{{\cal D}}
\def\bra{{\langle}}
\def\ket{{\rangle}}
\def\V{\mbox{\bf V}}
\def\T#1{\mbox{\bf T}_{#1}}

\def\GeV{\unskip\,\mathrm{GeV}}
\def\Vone{{ AutoDipole\ }}
\def\Vtwo{{ Ref.~\cite{Dittmaier:2008uj}\ }}
\newcommand{\Pt}{t}

\newcommand{\Pg}{g}

\def\slash#1{\rlap{\hbox{$\mskip 1 mu /$}}#1} 

\newcommand{\ep}{\epsilon}
\newcommand{\cd}{\cdot}

\newcommand{\al}{\alpha_{s}}

\begin{document}

\begin{titlepage}
  \begin{flushright}
    DESY-09-194\\
    HU-EP-09/55\\
    SFB/CPP-09-107\\[0.5cm]
    November 2009           
  \end{flushright}
  
  \vspace*{1.3cm}

  \begin{center}
    {\Large\bf AutoDipole \\[1ex] -- Automated generation of dipole subtraction 
      terms --}\\[1.5cm]

    {\large K. Hasegawa$^{\, a}$, S. Moch$^{\, b}$ and P. Uwer$^{\, a}$}
    \vspace*{1.2cm}

    {\it $^a$ Institut f\"ur Physik, Humboldt-Universit\"at zu Berlin,\\[0.1cm]
      D-10099 Berlin, Germany}\\[0.5cm]

    {\it $^b$Deutsches Elektronensynchrotron DESY \\[0.1cm]
      Platanenallee 6, D-15738 Zeuthen, Germany}\\[1.8cm]

    {\large\bf Abstract}\\[0.2cm]
    \parbox{0.8\textwidth}{
      We present an automated generation of the subtraction terms for 
      next-to-leading order QCD calculations in the Catani-Seymour dipole formalism.
      For a given scattering process with $n$ external particles 
      our Mathematica package generates all dipole terms, allowing for both massless and massive dipoles.
      The numerical evaluation of the subtraction terms proceeds with 
      MadGraph, which provides Fortran code for the necessary scattering amplitudes. 
      Checks of the numerical stability are discussed.
    }
  \end{center}
  \vfill
\end{titlepage}

%
%
\section*{Program summary}
%
%

{\it Title of program:} AutoDipole \\[2mm]
{\it Version:} 1.2.3 \\[2mm]
{\it Catalogue number:} \\[2mm]
{\it Program summary URL:} {\tt http://www-zeuthen.desy.de/\~{}moch/autodipole/} or\\
\hspace*{42mm}             {\tt http://www.physik.hu-berlin.de/pep/tools/}\\[2mm]
{\it E-mail:}               {\tt kouhei.hasegawa@physik.hu-berlin.de},\\
\hspace*{13mm}              {\tt sven-olaf.moch@desy.de},\\
\hspace*{13mm}              {\tt peter.uwer@physik.hu-berlin.de} \\[2mm]
{\it License:} ---\\[2mm]
{\it Computers:} Computers running Mathematica (version 7.0). \\[2mm]
{\it Operating system:} The package should work on every Linux
  system supported by Mathematica. Detailed tests have been performed
  on {\tt Scientific Linux} as supproted by DESY and CERN and on 
  {\tt openSUSE} and {\tt Debian}. \\[2mm]
{\it Program language:} Mathematica and Fortran. \\[2mm]
{\it Memory required to execute:} Depending on the complexity of the problem, 
        recommended at least 128 MB RAM.\\[2mm]
{\it Other programs called:} MadGraph (stand-alone version MG\_ME\_SA\_V4.4.30) and HELAS \\[2mm]
{\it External files needed:} MadGraph (including HELAS library) available under \\
\hspace*{37mm}              {\tt{http://madgraph.hep.uiuc.edu/}} or \\ 
\hspace*{37mm}              {\tt{http://madgraph.phys.ucl.ac.be/}} or \\ 
\hspace*{37mm}              {\tt{http://madgraph.roma2.infn.it/}} \\[2mm]
{\it Keywords:} QCD, NLO computations, dipole formalism, Catani-Seymour subtraction. \\[2mm]
{\it Nature of the physical problem: } 
        Computation of next-to-leading order QCD corrections to scattering cross sections, 
        regularization of real emission contributions. \\[2mm]
{\it Method of solution:} 
      Catani-Seymour subtraction method for massless and 
      massive partons~\cite{Catani:1996vz,Catani:2002hc}; 
      Numerical evaluation of subtracted matrix elements interfaced to 
      MadGraph~\cite{Stelzer:1994ta,Maltoni:2002qb,Alwall:2007st} (stand-alone version) 
      using helicity amplitudes and the HELAS library~\cite{Hagiwara:1990dw,Murayama:1992gi}
      (contained in MadGraph).
\\[2mm]
{\it Restrictions on complexity of the problem:} 
      Limitations of MadGraph are inherited.
\\[2mm]
{\it Typical running time:} Dependent on the complexity of the problem with 
typical run times of the order of minutes.

\newpage

%
%
\section{Introduction}
\label{sec:intro}
%
%
The Large Hadron Collider (LHC) allows us to explore an energy regime far beyond what has been accessible in direct measurements up to now. 
Operating at the TeV scale it will shed light on the mechanism of electroweak symmetry breaking. 
The large center-of-mass energy allows the production of new heavy
particles --- if they exist. 
It is expected that the results obtained from the LHC will
significantly influence our future understanding of nature. 
The high potential of the LHC in making an important step forward towards a
deeper understanding of elementary particle physics comes not for free. 
Experimentally as well theoretically the LHC is a very
challenging experiment. The large multiplicity in the individual event 
together with pile-up and the complications due to the underlying
event make the experimental analysis at the LHC highly non-trivial. 
The separation of
known Standard Model (SM) physics from new physics is very demanding,
in particular since in many cases the new physics signals are
overwhelmed by the large SM backgrounds. 
Sophisticated methods have been devised in the past to cope with this
situation within the experiments with respect to theoretical predictions.

Colliding protons at LHC interact primarily through strong interactions
and Quantum Chromodynamics (QCD) plays an important role in this context. 
It is well known that perturbative QCD predictions are in general plagued 
by a large (residual) renormalization and factorization scale dependence. 
It is not rare that predictions for cross sections
change by 100\% when the scales are varied in a
reasonable range. To reduce the large scale uncertainty
next-to-leading order (NLO) calculations are required, which 
generally consist of two ingredients: 
one contribution from the virtual corrections and a second contribution 
from real emission of one additional parton. 
With respect to the virtual contributions much progress has been achieved
recently, although the evaluation of one-loop amplitudes for ``large'' 
multiplicities ($2\to3$, $2\to4$) is still a highly non-trivial enterprise.
Fortunately, as far as the real corrections are concerned the situation is
much better and efficient methods exist for the numerical evaluation 
of the required matrix elements.
However, the integral over the phase space leads to soft and
collinear  singularities, which we denote here generically as infrared (IR) divergencies.
In combination with the virtual corrections, all IR cancel between the 
two contributions for the physical observables of interest~\cite{Bloch:1937pw,Kinoshita:1962ur,Lee:1964is}.

General algorithms, typically classified either as slicing or subtraction methods, 
are available for the extraction of soft and collinear singularities encountered in the real corrections.
In both cases one makes use of the universal behavior of QCD amplitudes for soft and collinear configurations.
IR singularities are nowadays usually regulated within dimensional
regularization, so the same regulator has to be applied to the real
corrections in a way that numerical integration over the phase space 
in 4 dimensions is possible in the end.

In slicing methods the idea is to separate the phase space into
{\it resolved} and {\it unresolved} contributions~\cite{Giele:1991vf,Giele:1993dj, Keller:1998tf,Harris:2001sx}. 
Unresolved regions at NLO are those where one parton becomes soft or two become collinear. 
In these regions the matrix elements are approximated using the QCD factorization theorems. 
After this simplification the unresolved regions can be integrated analytically in $d=4-2\ep$ dimensions 
and the emerging singularities cancel analytically against the corresponding ones in the virtual corrections. 
Resolved regions on the other hand exclude all IR singularities by definition
and can be integrated numerically in 4 dimensions. 
A one-dimensional illustration of the slicing approach is shown below, 
\begin{eqnarray}
  \int_0^1 {f(x)\over x^{1-\epsilon}}\, dx
  &\approx& f(0)\int_0^\delta {1\over x^{1-\epsilon}}\, dx 
  + \int_\delta^1 {f(x)\over x}\, dx + O(\epsilon)\nonumber \\
  &=& {1\over \epsilon}f(0)+\ln(\delta)f(0)  
  + \int_\delta^1 {f(x)\over x}\, dx + O(\epsilon)
  \, ,
  \label{eq:slicing}
\end{eqnarray}
where $f(x)$ is an arbitrary function, which is regular at $x=0$.
It is a drawback of the slicing method that partial results exhibit 
a logarithmic dependence on the cut $\delta$ which separates resolved from unresolved configurations. 
This logarithm --- which cancels when the two parts (resolved and unresolved) are combined --- 
is manifest in the analytic integration of the unresolved terms.
However, for the resolved terms the logarithmic dependence arises from the numerical
phase space integration. 
Since the matrix elements in the unresolved phase space regions are only approximate one tends
to make the respective region around singular configurations as small as possible. 
This procedure, though, would result in large numerical cancellations and a loss of accuracy in
the sum of the resolved and unresolved parts compared to the accuracy reached in the numerical integration.  
In practice this requires a compromise between the quality of the approximation 
and the numerical effort to achieve a certain precision in the sum of both,
resolved and unresolved contributions.

Subtraction methods make use of our knowledge about QCD factorization 
in the soft and collinear limits to construct suitable ``counter-terms''. 
These have to match pointwise all singularities in the real-emission matrix elements 
and, at the same time, should be simple enough to be integrated analytically in $d=4-2\ep$ dimensions 
over the entire phase space~\cite{Catani:1996vz,Frixione:1995ms,Catani:2002hc}.
A one-dimensional example is shown in \Eq{eq:subtraction}.
\begin{eqnarray}
  \int_0^1 {f(x)\over x^{1-\epsilon}}\, dx &=&
  \int_0^1 {f(x)-f(0)\over x}\, dx  
  + f(0) \int_0^1 {1\over x^{1-\epsilon}}\, dx + O(\epsilon)\nonumber\\
  &=&
  \int_0^1 {f(x)-f(0)\over x}\, dx  
  + {1\over \epsilon}f(0) + O(\epsilon)
  \, . 
  \label{eq:subtraction}
\end{eqnarray}
Here the cancellation of the singularity takes place at the integrand
level --- i.e. the divergencies cancel pointwise --- and the
integrals are easier to evaluate numerically. 
However, care must be taken with respect to the numerical accuracy.
Deep in the singular regions the individual contributions become arbitrary
large and, due to limited numerical precision of floating point arithmetic, 
their cancellation might be incomplete leading to potentially wrong results. 
While our one-dimensional example in \Eq{eq:subtraction} admits the construction of
a suitable subtraction term in a  straightforward manner this no longer true when  
considering complicated scattering amplitudes. 
Fortunately, this formidable problem has been solved with the 
Catani-Seymour dipole formalism~\cite{Catani:1996vz,Catani:2002hc}. 
(Similar algorithms are presented in Refs.~\cite{Frixione:1995ms,Phaf:2001gc}.) 
Based on the factorization of soft and collinear singularities in an $\mbox{SU}(N)$ gauge theory all 
subtraction terms are constructed from universal functions and process specific amplitudes. 
It turns out, though, that for complicated processes the complete subtraction term is a sum over many different
contributions --- tedious to derive by hand. 
On the other hand, the method is completely algorithmic, thus an automation is feasible. 
This is a timely problem and its solution is the aim of the present work~\cite{Hasegawa:2008ae,Hasegawa:2009zz}. 
We note that it has recently also been addressed by other 
groups~\cite{Gleisberg:2007md,Seymour:2008mu,Frederix:2008hu,Czakon:2009ss},
see also \cite{Frederix:2009yq}.

In passing let us briefly mention that the attractive features of the subtraction approach 
(universality of counter-terms and numerical stability) 
have led to extensions of the formalism to next-to-next-to-leading order (NNLO), 
see e.g. \cite{GehrmannDeRidder:2004tv,Weinzierl:2003fx,Frixione:2004is,Somogyi:2005xz}. 
Presently, the proposed schemes at NNLO apply to processes without colored partons in the initial state and an arbitrary number 
of massless particles (colored or colorless) in the final state. 
The necessary counter-terms are either derived from so-called antenna functions~\cite{GehrmannDeRidder:2005cm}
or alternatively defined as universal counter-terms based on QCD factorization
in the various soft and collinear limits of singly- and double-unresolved
parton configurations~\cite{Somogyi:2006db,Somogyi:2006da,Aglietti:2008fe,Bolzoni:2009ye}.
Ongoing work at NNLO is concerned with extensions to colored partons in the initial state as needed for the LHC.

The outline of the paper is as follows. 
In Section~\ref{sec:review} we review the general features of the
Catani-Seymour algorithm~\cite{Catani:1996vz,Catani:2002hc} and in Section~\ref{sec:package} 
we describe the details of its implementation in the AutoDipole package. 
Particular emphasis is put on the details of the numerical evaluation of
subtraction terms via an interface to MadGraph~\cite{Stelzer:1994ta,Maltoni:2002qb,Alwall:2007st}, 
which uses the HELAS library~\cite{Hagiwara:1990dw,Murayama:1992gi} 
for the computation of helicity amplitudes. 
Section~\ref{sec:userguide} illustrates with a few examples of how to use AutoDipole in practice. 
The examples serve also as a non-trivial
cross check of the implementation when compared with existing results
from the literature. 
Finally, we conclude in Section~\ref{sec:conclusions} and list some technical details 
of the implementation and the comparison with the literature 
in Appendices~\ref{sec:appA}--\ref{sec:appC}.

%
%
\section{Review of the Catani-Seymour subtraction formalism}
\label{sec:review}
%
%
We consider a generic scattering process for the production of an $n$-parton final state.  
At NLO accuracy in QCD the corresponding cross section may be written as:
\begin{equation}
  \sigma_{\NLO} = \sigma_{\LO} + \delta \sigma_{\NLO}
\, ,
\end{equation}
where $\sigma_{\LO}$ denotes the Born contribution at leading order (LO) 
and the genuine NLO correction $\delta \sigma_{\NLO}$  receives contributions from three different sources,
\begin{equation}
  \label{eq:nlo-separation}
  \delta \sigma_{\NLO} =  \int_n d\sigma_{\virt} +  
  \int_{n+1} d\sigma_{\real} +
    \int dx\int_n d\sigma_{\fac} 
\, .
\end{equation}
By $d\sigma_{\virt}$ we denote here the contributions from the virtual corrections 
and by $d\sigma_{\real}$ the ones from the real emission of one additional parton.
For hadrons in the initial state there is a third
contribution $d\sigma_{\fac}$ due to the factorization of initial
state singularities. 
The subscript on the integral signs in \Eq{eq:nlo-separation} indicates the dimensionality of the phase space:
The virtual corrections are integrated over an $n$-parton phase space
while the real emissions are integrated over an $(n+1)$-parton phase space.
All three contributions are individually divergent and thus require regularization 
in intermediate steps until the divergencies are canceled. 
In the Catani-Seymour formalism the expression for $\delta \sigma_{\NLO}$ 
in \Eq{eq:nlo-separation} is rewritten schematically:
\begin{equation}
\label{eq:master}
  \delta \sigma_{\NLO} =  
   \int_{n+1} (d\sigma_{\real} + dA )
  + \int_n (d\sigma_{\virt} + \int_1 dA') 
  + \int dx \int_n (d\sigma_{\fac}+dA'') 
  \, ,
\end{equation}
with 
\begin{equation}
  0 =  
  \int_{n+1} dA 
 + \int_n \int_1 dA' 
 + \int dx \int_n dA''
 \, .
\end{equation}
The expressions $dA$, $dA'$ and $dA''$ are defined such that they
render the individual pieces in \Eq{eq:master} finite. 
In the Catani-Seymour formalism, they correspond 
to the sum of all dipoles ($dA$), the integrated dipoles ($\int_1 dA'$), ${\mbox{\bf I}}$-term, 
and the terms arising from mass factorization ($dA''$) 
including the so-called ${\mbox{\bf P}}$-, ${\mbox{\bf K}}$- and ${\mbox{\bf H}}$-terms.

The explicit form of $dA$ is constructed from the knowledge about the soft and
collinear factorization of QCD amplitudes, which exhibit a simple factorization for collinear configurations.
The expression for $dA$ in \Eq{eq:master} is obtained as a sum over potentially collinear partons 
since the factorization for soft configurations can be derived from the collinear behavior of the amplitudes.
For soft singularities however only color-ordered amplitudes exhibit a simple
factorization, which implies in general non-trivial color correlations at the level of the squared matrix elements.
These color correlations are reflected by the reference to an additional (so-called) spectator parton. 
The subtraction $dA$ is thus written as a sum of individual dipoles in the following form:
\begin{equation}
  \label{eq:master1}
  dA = \sum \Dipole(i,j;k)
\, ,
\end{equation}
where the sum runs over all colored partons in scattering process and 
the possible configurations for $\{i,j;k\}$ are determined from the
real corrections for which the subtraction term is constructed. 
The generic form of the dipoles is given by
\begin{equation}
  \label{eq:master2}
 \Dipole(i,j;k) = d_{ij}\, \times \, 
 \bra 1,\ldots \widetilde{ij},\ldots,\tilde k,\ldots,n | \V_{ij,k} 
 | 1,\ldots \tilde{ij},\ldots,\tilde k,\ldots,n\ket
 \, .
\end{equation}
The singular behavior is contained in the pre-factor $d_{ij}$ which is 
essentially the intermediate propagator before the splitting into $i+j$.
The bra ($\bra \dots |$)and ket ($| \dots \ket$) notation for the amplitude 
is used since it appears as a vector in color and spin space. 
Accordingly $\V_{ij,k}$ acts as an operator 
in color and spin space and introduces the non-trivial color and 
spin correlations mentioned above.

As a concrete example we display the expression for $\Dipole(g_i,g_j,k)$ 
with all three partons in the final state, i.e. 
\begin{equation}
  \label{eq:dipggk} 
  \Dipole(g_i,g_j,k) = -{1\over 2 p_i \cdot p_j}
  \bra 1,\ldots \widetilde{ij},\ldots,\tilde k,\ldots,n |
  {\T{k}\cdot\T{ij}\over \T{ij}^2} V_{g_ig_j,k} 
  | 1,\ldots \tilde{ij},\ldots,\tilde k,\ldots,n\ket
  \, .
\end{equation}
$\Dipole(g_i,g_j,k)$ describes the situation of a collinear splitting of 
the gluons $i$ and $j$ (emitter pair) in the presence of a spectator parton $k$.
Here, the $\T{i}$ are the color charge operators depending on the partons being in
the fundamental or adjoint representation of the color $\mbox{SU}(3)$ 
(for details we refer to \cite{Catani:1996vz,Catani:2002hc}). 
The function $V_{g_ig_j,k}$ is given by
\begin{eqnarray}
  \label{eq:vggk}
V_{g_ig_j,k}^{\mu\nu} &=& \bra\mu|V_{g_ig_j,k}|\nu\ket \nonumber \\ 
&=& 16 \pi \al \ \mu^{2 \ep}\ C_A\ \biggl[
-g_{\mu \nu} \biggl(   \frac{1}{1 - z_{i} (1 - y_{ij,k})} + \frac{1}{1 - z_{j} 
(1 - y_{ij,k})} - 2 \biggr)  \nonumber \\
& & + (1 - \ep)\frac{1}{p_{i} \cdot p_{j}}(z_{i}p_{i}^{\mu} - z_{j}p_{j}^{\mu})(z_{i}p_{i}^{\nu} 
- z_{j}p_{j}^{\nu}) \biggr]
\, ,
\end{eqnarray}
where $z_{i}$ and $y_{ij,k}$ are some functions of the Lorentz scalars $s_{ij}, s_{ik},$
and $s_{jk}$ and $\al$ is the strong coupling. The quantity, 
\begin{eqnarray}
\label{eq:clbs}
\bra 1,\ldots \widetilde{ij},\ldots,\tilde k,\ldots,n |
  \T{k}\cdot\T{ij}
  | 1,\ldots \tilde{ij},\ldots,\tilde k,\ldots,n\ket,
\end{eqnarray}
is called the color linked Born amplitude squared (CLBS). 
The amplitude ($| \dots \rangle$) is deduced from the real emission by the factorization of one splitting. 
The color factor is extended by two color operator insertion, $\T{k} \cdot \T{ij}$. 
The CLBS has the reduced kinematics as input. In the example
the original momenta $(p_{i},p_{j},p_{k})$ are reduced to $(\tilde{p}_{ij}, \tilde{p}_{k})$ 
where for example, the $\tilde{p}_{ij}$ is defined as
\begin{eqnarray}
\label{eq:reducedkin}
\tilde{p}_{ij}^{\mu} = p_{i}^{\mu} + p_{j}^{\mu} - \frac{y_{ij,k}}{1 - y_{ij,k}}p_{k}^{\mu}.
\end{eqnarray}
Generally, $(n+3)$ original momenta are reduced to the $(n+2)$  momenta in each dipole term. 
One distinguished feature of the dipole subtraction is that the reduced 
kinematics satisfy the on-shell conditions and momentum conservation, which makes
it possible to evaluate the reduced Born amplitude by existing codes for
LO calculations. 
In the gluon emitter case like the example above, the second term in
the square bracket in \Eq{eq:vggk} introduces spin correlations of the gluon. 
In that case we need also to evaluate the CLBS where the helicity of the emitter gluon 
in the amplitude is different from the one in the conjugate. 
In the next section we explain how the three main ingredients, 
the generation of the dipole terms, the CLBS, 
and the dipoles with a gluon emitter, are implemented in our package. 

The other two functions $\int_1 dA'$ and $dA''$ in \Eq{eq:master} are easily explained.
The first one ($\int_1 dA'$) denotes the integrated dipoles which cancel the
corresponding IR divergencies in the virtual contributions at NLO.
They can be constructed from universal insertion operators ${\mbox{\bf I}}$,
to be sandwiched between the bra ($\bra \dots |$)and ket ($| \dots \ket$)
amplitudes of the corresponding Born process, i.e. in the reduced kinematics with $(n+2)$  momenta. 
The expressions for the ${\mbox{\bf I}}$-terms depend on the parton type,
e.g. for massless partons we have
\begin{eqnarray}
  \label{eq:Iterm}
  {\mbox{\bf I}}(\{p\},\ep) &=& - {\al \over 2 \pi} {1 \over \Gamma(1-\ep)} 
  \sum_{i} {1 \over \T{ij}^2}\ {\cal V}_i(\ep) \sum_{j \neq i} \T{i} \cd \T{j} 
  \left( {4 \pi \mu^2 \over 2 p_i \cd p_j} \right)^\ep\, ,
\end{eqnarray}
where the sum runs over all parton momenta $\{p\}$ in the Born kinematics (corresponding also to the NLO virtual corrections).
The poles in dimensional regularization are contained in the function ${\cal V}_i$,
that is ${\cal V}_i(\ep) \sim 1/\ep^2$ and $\sim 1/\ep$ for massless partons~\cite{Catani:1996vz}.
For massive partons \cite{Catani:2002hc}, the corresponding function ${\cal V}_i(\ep,m_i,m_j)$ 
contains single poles ${\cal V}_i(\ep,m_i,m_j) \sim 1/\ep$ from soft gluons 
and the ${\cal V}_i$ depend logarithmically on the parton masses $m_i$ which screen the collinear singularity, 
see also \cite{Phaf:2001gc,Mitov:2006xs}. 
The formulation suggested in \cite{Catani:2002hc} (and used in AutoDipole)
allows for a smooth interpolation in the limit $m_i \to 0$.
The color operator insertions, $\T{i} \cdot \T{j}$,   
lead again to non-trivial color correlations and require the evaluation 
of the same CLBS in \Eq{eq:clbs} with the phase space of $n$ final state partons.

The last term in \Eq{eq:master} ($dA''$) abbreviates the so-called 
${\mbox{\bf P}}$-, ${\mbox{\bf K}}$- and ${\mbox{\bf H}}$-terms. 
They are needed for mass factorization of initial state singularities in hadron collisions 
to be absorbed in renormalized parton distributions and, likewise, 
also for the final-state singularities in processes with identified hadrons
giving rise to scale dependence of the fragmentation functions.
For instance in the case of initial state divergencies the 
${\mbox{\bf P}}$-operator takes the following form
\begin{eqnarray}
  \label{eq:Pterm}
  {\mbox{\bf P}}^{a,b}(\{p\},x,\mu_F^2) &=& {\al \over 2 \pi}\ P^{ab}(x) \ 
  {1 \over \T{b}^2} \sum_{i \neq b} \T{i} \cd \T{b} 
  \ln {\mu_F^2 \over 2 xp_a \cd p_i} \, ,
\end{eqnarray}
where $\mu_F$ is the factorization scale and $P^{ab}$ are the standard (space-like) LO splitting functions 
containing the well-known '+'-distributions in their diagonal components.
Like the ${\mbox{\bf I}}$-operators in \Eq{eq:Iterm}, 
the ${\mbox{\bf P}}$-operators act on the amplitudes of the corresponding Born process, 
leading to color correlations and giving rise to the CLBS as in \Eq{eq:clbs}.
Here the dependence of ${\mbox{\bf P}}$ in \Eq{eq:Pterm} on the parton momentum fraction $x$ leads to convolutions 
when integrated with the Born squared matrix elements or CLBS, see \Eq{eq:master}, 
and the implementation needs a prescription for '+'-distributions (see
e.g. Eq.~(B.26) in \cite{Floratos:1981hs}).

Similar definitions as in \Eq{eq:Pterm} hold for the ${\mbox{\bf K}}$- and ${\mbox{\bf H}}$-operators. 
They also contain distributions $1/(1-x)_+$ that are singular in the collinear limit and 
parametrize the factorization scheme dependence, ${\mbox{\bf K}}$ for factorization
in the initial state and ${\mbox{\bf H}}$ in the final state.
These remarks conclude our brief review of the Catani-Seymour dipole
formalism. For details the reader is referred to the original literature~\cite{Catani:1996vz,Catani:2002hc}.

%
%
\section{The AutoDipole package}
\label{sec:package}
%
%
The AutoDipole package constructs all necessary subtraction terms of the Catani-Seymour
formalism~\cite{Catani:1996vz,Catani:2002hc} for a given
scattering process with $n$ colored partons in a fully automatic manner.  
It can handle both massless and massive partons and it also allows 
for additional (non-colored) SM particles in the scattering process, 
e.g. couplings of quarks to $\gamma$, $Z$-, $W^\pm$-bosons and so on.

Let us briefly sketch the details of our implementation of the dipole
subtraction formalism, because there is a large freedom in the way how this can be performed. 
For instance, in Refs.~\cite{Dittmaier:2007wz,Dittmaier:2008uj} the implementation
was realized in form of two independent C/C++ libraries providing all the
necessary functions to evaluate the dipole terms. 
As a slight disadvantage of this approach the produced code is non-local 
and that there is some redundancy in the calculation. 
In the present work we follow a different strategy. 
The main idea here is to have a code generator which will produce an optimized flat code.
To realize this we have constructed a Mathematica program  which acts as such a generator 
and its output is interfaced with MadGraph~\cite{Stelzer:1994ta,Maltoni:2002qb,Alwall:2007st}.
To that end we found the stand-alone version of MadGraph suitable and for the
numerical evaluation of all subtraction terms, we use helicity amplitudes
provided by MadGraph (which are based on the HELAS library~\cite{Hagiwara:1990dw,Murayama:1992gi}).

\subsection{Structure of the code}
\label{sec:code}

\begin{figure}[tb]
  \begin{center}
    \leavevmode
     \includegraphics[width=0.6\textwidth]{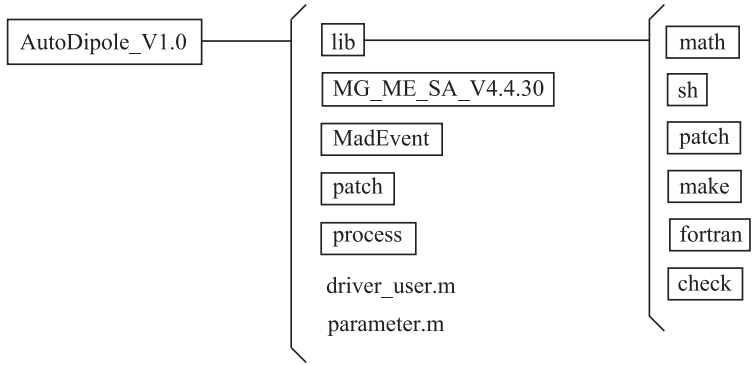}
    \caption{\small
      The directory structure of the AutoDipole package.}
    \label{fig:fig1}
  \end{center}
\end{figure}
The complete layout of the code of the AutoDipole package is displayed in Figure~\ref{fig:fig1}.
The major part of the source code in the package (all contained in the directory {\tt lib}) is written in Mathematica. 
This includes all algorithms for the generation of the dipole terms as well as 
the evaluation of the ${\mbox{\bf I}}$-, ${\mbox{\bf P}}$- and ${\mbox{\bf K}}$-operator 
insertions sandwiched between the Born amplitudes in the appropriate kinematics.
The use of a computer algebra program like Mathematica has the clear advantage here that
all terms are accessible to symbolic manipulation.
This feature is very useful for general studies of the IR behavior of scattering 
amplitudes.

\begin{figure}[tb]
  \begin{center}
    \leavevmode
     \includegraphics[width=0.49\textwidth]{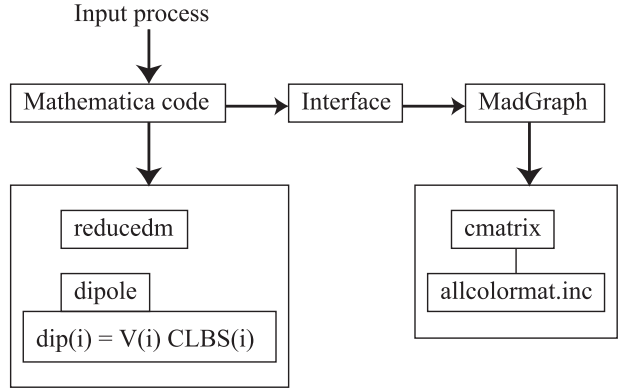}
    \caption{\small
      The flowchart of the execution of AutoDipole.}
    \label{fig:flowchart}
  \end{center}
\end{figure}

In using AutoDipole 
we are primarily interested in the generation of Fortran code for all subtraction terms of \Eq{eq:master}.
The essential parts of the flowchart are displayed in Figure~\ref{fig:flowchart}.
From a given input process, the Mathematica code generates all dipole terms. 
The important files here are {\tt dipole.f}, {\tt reducedm.f}, and the shell script 
for the interface with MadGraph, where {\tt dipole.f} contains all dipole terms except for 
the reduced kinematics and the CLBS. 
The routine {\tt reducedm.f} calculates the reduced kinematics of each dipole
term as defined in \Eq{eq:reducedkin}.
The subsequent run of a patched version of MadGraph (contained in the directory {\tt patch}) via the interface
produces the files, {\tt cmatrix.f} and {\tt allcolormat.inc}. 
The routine {\tt cmatrix.f} evaluates all CLBS as in \Eq{eq:clbs}. 
It includes the file {\tt allcolormat.inc}, which contains all extended color matrices 
by the two color operator insertions. 
We use the latest version of MadGraph (MG\_ME\_SA\_V4.4.30).
All the process specific Fortran files mentioned above are stored in a new directory under {\tt process}, see Figure~\ref{fig:fig1}.

In short, the evaluation of the generated Fortran code for the dipole terms
proceeds along the following chain, see Figure~\ref{fig:flowchart}.
The routine {\tt reducedm.f} receives a given phase space point as input
and calculates the reduced kinematics of each dipole term.
Next, the routine {\tt cmatrix.f} (with the reduced kinematics as input) 
computes the CLBS of each dipole term together with the extended 
color matrices in {\tt allcolormat.inc}.
Finally the routine {\tt dipole.f} returns the values of all dipole terms 
having received the reduced kinematics and the CLBS as input.
All this is done automatically when the user executes the command {\tt ./runD} 
(see Sections~\ref{sec:short-example} and~\ref{sec:use}) to calculate the subtraction term for a specific phase space point.

The algorithm of dipole term generation by the Mathematica code is shown in Section~\ref{sec:crea}.
A detailed account of the calculation of the CLBS is given in Section~\ref{sec:colo} 
and some aspects of the spin correlations for the particular case of a gluon emitter 
are presented in Section~\ref{sec:gluo}.
The check of IR safety is explained in Section~\ref{sec:safe}.
An explanation of how to run AutoDipole is deferred to Section~\ref{sec:userguide}.

\subsection{Algorithm of dipole generation}
\label{sec:crea}
Let us start with a short description of the Mathematica code generating all dipole terms.
This procedure makes use of the following three steps:
\begin{figure}[htbp]
  \begin{center}
    \leavevmode
     \includegraphics[width=0.49\textwidth]{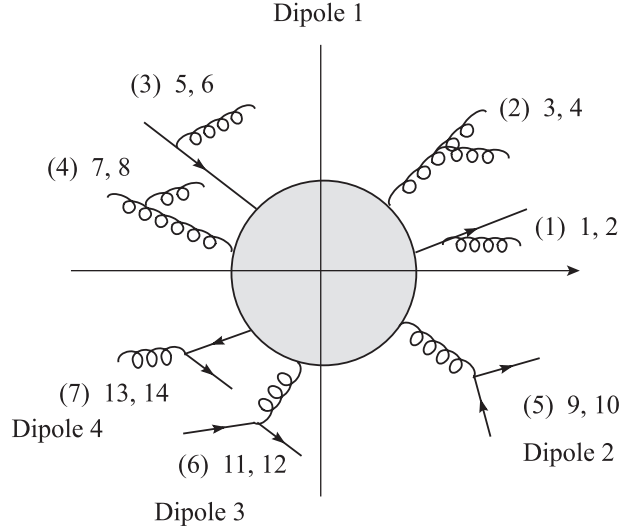}
    \caption{\small
      The four categories of dipoles (Dipole 1, ..., 4), the seven possible
      splittings and their order of creation in AutoDipole, 
      see Table~\ref{tab:creation-order-dip} for details.}
    \label{fig:layout-dip}
  \end{center}
\end{figure}

\bigskip

{\bf 1.} Choose all possible emitter pairs from the external legs. \\
For a given real emission $2 \to (n+1)$-particle scattering process 
we abbreviate the set of initial (final) state partons collectively as
$\{\mbox{initial}\}$ ($\{\mbox{final}\}$), 
i.e. the scattering reaction reads generically 
\begin{equation}
  \label{eq:def-infin}
  \{\mbox{initial}\} \,\rightarrow\, \{\mbox{final}\}
  \, .
\end{equation}
Then, the first step in the construction of the dipoles is the choice of the emitter
pair, that is the root of the splitting of the quarks and gluons.
This also specifies the kind of splitting. 
At NLO in QCD, there exist seven possible splittings, which we group into four classes and 
we enumerate these types of dipoles accordingly, Dipole 1, ..., and Dipole 4. 
The ordering is shown in Figure~\ref{fig:layout-dip}.
Due to the factorization of the splitting each dipole is associated to a certain reduced Born process 
and the following equations show schematically the operations performed on the
sets of partons in Eq.~(\ref{eq:def-infin}) in order to construct the corresponding reduced Born process,
\begin{eqnarray}
  \label{eq:dipole-cat1}
\mbox{Dipole\ 1:}& \{\mbox{initial}\} \phantom{-f+g} &\rightarrow\, \{\mbox{final}\}  -g
\, , \\
  \label{eq:dipole-cat2}
\mbox{Dipole\ 2:}& \{\mbox{initial}\} \phantom{-f+g} &\rightarrow\, \{\mbox{final}\} -f\bar{f}+g 
\, , \\
  \label{eq:dipole-cat3}
\mbox{Dipole\ 3:}& \{\mbox{initial}\} -f+g &\rightarrow\, \{\mbox{final}\} -f 
\, , \\
  \label{eq:dipole-cat4}
\mbox{Dipole\ 4:}& \{\mbox{initial}\} -g+\bar{f} &\rightarrow\, \{\mbox{final}\} -f \, ,
\end{eqnarray}
where $g$ is a gluon and $f(\bar{f})$ a quark (anti-quark).
The notation in \Eqs{eq:dipole-cat1}{eq:dipole-cat4} indicates which partons are removed from or added to the
respective sets $\{\mbox{initial}\}$ and $\{\mbox{final}\}$. 
The cases Dipole 3 and 4 can also occur with a reduced Born cross section where $f$ and $\bar{f}$ are exchanged.
The groups Dipole 2, 3 and 4 also need to account for the existence of various quark flavors, 
i.e. they exhibit manifest flavor dependence, see Table~\ref{tab:creation-order-dip} for details.

{\bf 2.} Choose all possible spectators for each emitter pair. \\
The spectator is one external field which is different from both fields
of the emitter pair and its choice is a purely combinatorial procedure. 
We use indices, $i, j,$ and $k$, for fields in a final state and,
respectively, indices, $a$ and $b$, for an initial state field. 
For a spectator in the final (initial) state denoted by $k$ ($b$), this condition means $k \not= i, j$ ($b \not= a$). 
It emerges from a special feature of the subtraction formalism namely 
that the squared matrix elements with the Casimir operator, $\bra \ \T{i}^{2} \ \ket$,
is expressed, due to color conservation, through CLBS $\bra \ \T{i}\cdot \T{k} \ \ket (i \not= k)$
which always includes one color dipole configuration in the part.

{\bf 3.} Construct the dipole terms from the chosen combinations of emitter and spectator. \\
The previous steps provide all such combinations as pairings of the type (emitter, spectator)$=(ij,k),(ij,b),(ai,k)$, and $(ai,b)$.
Each case corresponds to one dipole term, for which we use the short-hands 
$\mbox{D}_{ij,k}$, $\mbox{D}_{ij}^{a}$, $\mbox{D}^{ai}_{k}$, and $\mbox{D}^{ai,b}$, respectively.
Detailed information about all pairings of emitter and spectator as well as
the potential presence of a mass parameter is given in Table~\ref{tab:creation-order-dip} in Appendix~\ref{sec:appA}.  
For explicit expressions we refer to~\cite{Catani:1996vz,Catani:2002hc}. 

\bigskip

Here it is worth to stress a few points.
First of all, the chosen order for the generation has the advantage of
grouping together the reduced Born matrix elements. 
The CLBS in the same category exhibit the same Lorentz structure, 
only the extended color matrix is different.
This ordering leads to a block-diagonal structure in the evaluation of the
color correlations. It leads to flat Fortran code including higher readability.
Additionally, for the category Dipole 1 the Casimir operator $\T{ij}^2$ in the denominator 
always cancels against the same one in the dipole splitting function.
This cancellation leads to manifest simplifications.
Finally, in our set-up we have access to the symbolic expression and can easily identify or
extract partial subsets of dipole terms, if needed. 
For example we may want to discard the $t-\bar{t}$ (heavy-quark) splitting in
the category Dipole 2, 
because this splitting does not give rise to poles in $\ep$.
Our way of generating the dipoles in the package easily allows this selection. 
The user can discard dipoles of a specific category in the input file {\tt parameter.m} 
(see Section~\ref{sec:userguide}). 

\begin{figure}[htbp]
  \begin{center}
    \leavevmode
     \includegraphics[width=0.425\textwidth]{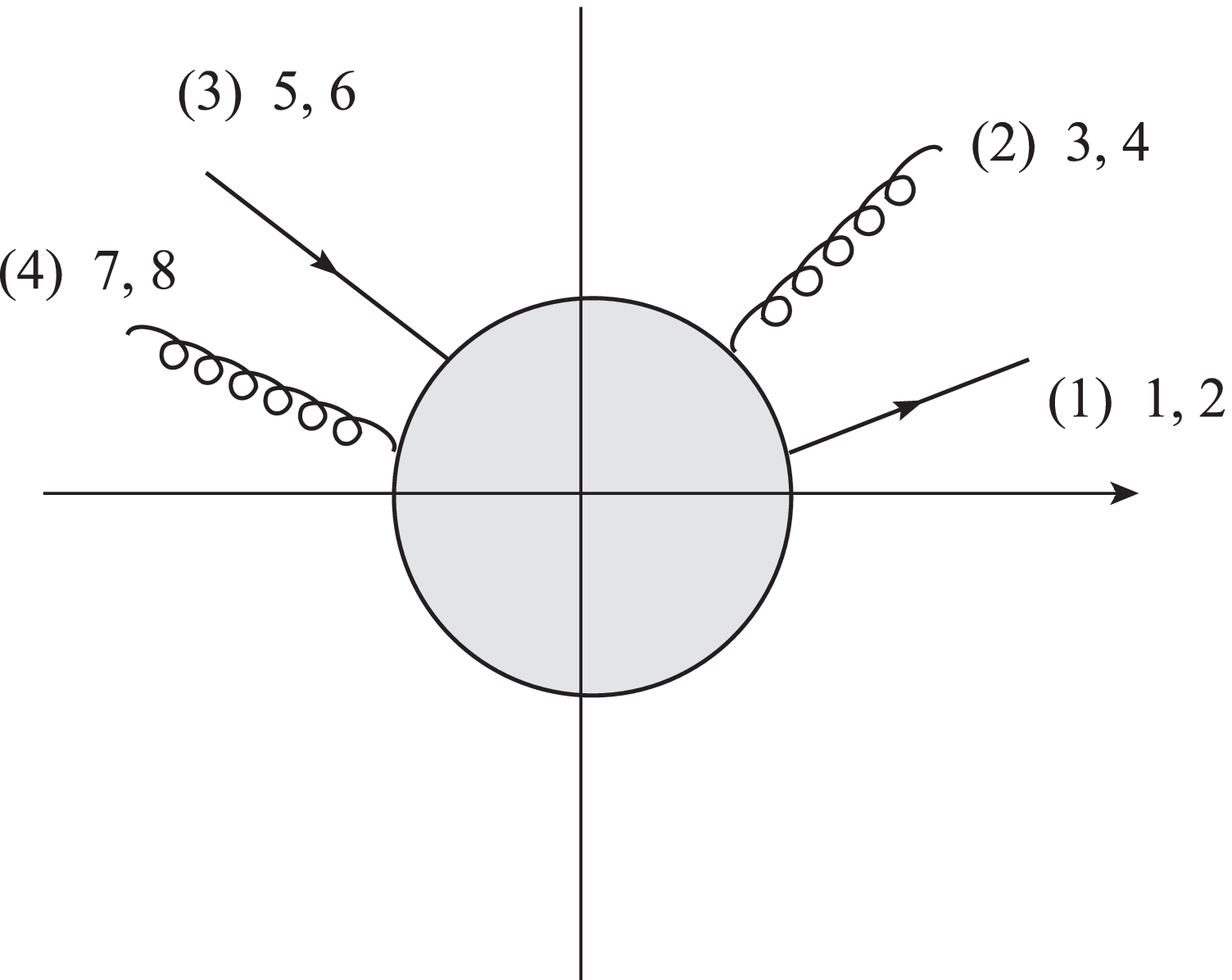}
     \hspace*{5.0mm}
     \includegraphics[width=0.40\textwidth]{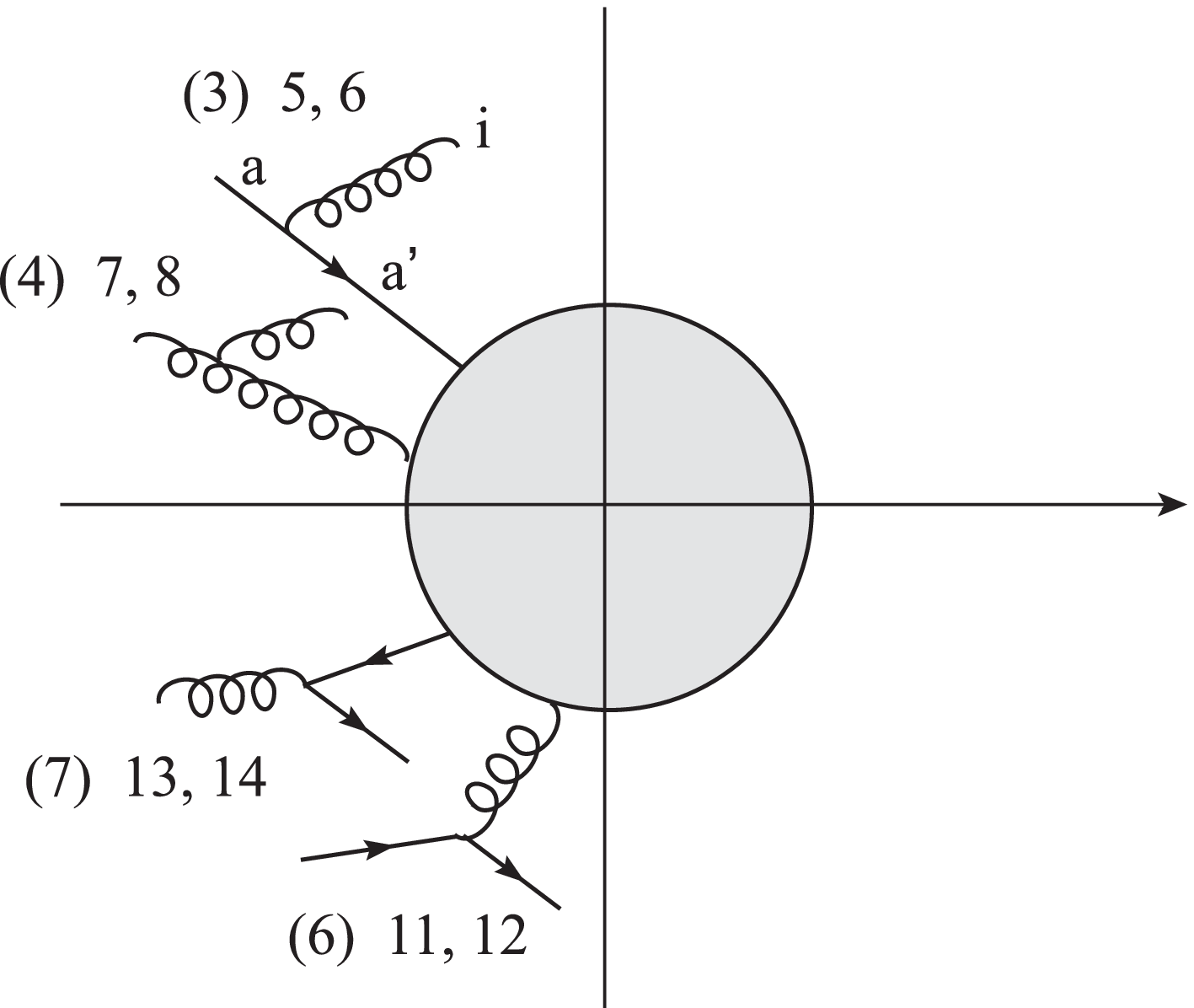}
    \caption{\small
      The order of creation of the integrated dipoles in AutoDipole with insertion of 
      the {\mbox{\bf I}}-operator (left) and 
      of the ${\mbox{\bf P}}$- and ${\mbox{\bf K}}$-operators (right),
      see Tables~\ref{tab:creation-order-I} and ~\ref{tab:creation-order-pandk} for details.}
    \label{fig:layout-IPK}
  \end{center}
\end{figure}

At this stage it remains to discuss the generation of all terms involving 
the ${\mbox{\bf I}}$-, ${\mbox{\bf P}}$- and ${\mbox{\bf K}}$-operators.
This is sketched in Figure~\ref{fig:layout-IPK}.
Given, that these operators originate from the phase space integral over unresolved parton of the dipole terms,
it is clear that we only need a subset of all previously generated terms.
From all reduced Born amplitudes generated for the regularization of the real emission contributions
(i.e. Figure~\ref{fig:layout-dip}) we need only the CLBS in the category Dipole 1 for all ${\mbox{\bf I}}$-,
${\mbox{\bf P}}$-, and ${\mbox{\bf K}}$-operator insertions. 
Hence, no new information is required and the Mathematica package assembles
the respective expressions (along with the color correlations, see Section~\ref{sec:colo}) in the Fortran
files {\tt Iterm.f} and {\tt PKterm.f}.

\subsection{The color linked Born amplitude squared}
\label{sec:colo}
The numerical evaluation of Born amplitudes (or their squares) is a routine
task for many publicly available packages designed for automated LO calculations. 
As announced above, 
we choose the stand-alone version of MadGraph~\cite{Stelzer:1994ta,Maltoni:2002qb,Alwall:2007st} for this purpose 
and the added feature needed is the evaluation of the color link operators, see e.g. \Eq{eq:dipggk}. 
This can be done by a patch and allows us to obtain all CLBS in an automatic way, see Figure~\ref{fig:flowchart}.

The generation of the amplitudes in MadGraph proceeds via Feynman diagrams 
and the color factors are separated from each diagram. 
During the evaluation everything is expressed in terms of generators of the fundamental representation 
of the color $\mbox{SU}(3)$.
A typical example is that the factor $f^{abc}$ of the gluon three-point vertex
is rewritten in terms of the fundamental generator $t^{a}_{ij}$ with the help of the identity,
\begin{eqnarray}
  f^{abc} = -2 i \bigl( \ \mbox{Tr}[t^{a}t^{b}t^{c}] - \mbox{Tr}[t^{c}t^{b}t^{a}] \ \bigr) 
  \, .
\end{eqnarray}
The color factors of each diagram are sorted in a unique order and they are expressed in a sum. 
When a specific term of a diagram is identical to one of the other diagrams,
both are combined as 
\begin{eqnarray}
  \mbox{M} = \sum_{a} C_{a} \mbox{J}_{a}
  \, ,
\end{eqnarray}
where $C_{a}$ denotes the independent color factors. 
Each $C_{a}$ has fundamental and adjoint color indices corresponding to the external quarks and the gluons, respectively.
$\mbox{J}_{a}$ is the joint amplitude, e.g. $\mbox{J}_{1}=+A_{1}-A_{3}+ \cdots$ where $A_{i}$ 
is the partial amplitude of $i$-th diagram (with the color factor stripped off).
The invariant matrix element squared is finally expressed in the form,
\begin{eqnarray}
  \label{eq:mesquared}
  |\mbox{M}|^{2} = \bigl(\vec{\mbox{J}}\bigr)^{\dagger} \ \mbox{CF} \ \vec{\mbox{J}}
  \, ,
\end{eqnarray}
where the color matrix $\mbox{CF}$ is defined as 
\begin{equation}
    \label{eq:cf}
(\mbox{CF})_{ab}=\sum_{\mbox{\scriptsize color}}C_{a}^{\ast} C_{b}.   
\end{equation}
For the CLBS we need to evaluate \Eq{eq:mesquared} with an insertion of
two additional color operators to the emitter and spectator legs.
This is precisely what our patch of MadGraph does (see Figure~\ref{fig:fig1}
and the directory {\tt lib/patch}).
The subroutines of MadGraph for the color factor calculations are well structured and the
original routines to add the color factors $t^{a}_{ij}$ and $f^{abc}$ can be applied to the
additional color insertions for the CLBS. 
The color algebra of the $\mbox{SU}(3)$ is performed numerically and 
the resulting extended color matrix $\mbox{CF}$ is written to the file {\tt allcolormat.inc}, see Figure~\ref{fig:flowchart}.

We have realized the two color operator insertions for all CLBS in an automatic way and 
we have also checked that MadGraph with our interface works for a set of rather involved processes.
As a simple example let us discuss the color insertions required for the process $g(a)g(b) \rightarrow u(1)\bar{u}(2)g(3)$. 
The reduced Born process $g(a)g(b) \rightarrow u(1)\bar{u}(2)$ has three diagrams and the
color factors are combined into two independent ones,
$(C_{1}, C_{2}) = ((t^{a} t^{b})_{12},(t^{b} t^{a})_{12})$. 
The components of the color matrix are given by the traces, 
$(\mbox{CF})_{11}=(\mbox{CF})_{22}=\mbox{Tr}[t^{b}t^{a}t^{a}t^{b}]$ and
$(\mbox{CF})_{12}=(\mbox{CF})_{21}=\mbox{Tr}[t^{b}t^{a}t^{b}t^{a}]$.
Then the color matrix is calculated as 
\begin{eqnarray}
  \mbox{CF} = \left( \begin{array}{rr}
      16/3 & -2/3  \\
      -2/3 & 16/3  \\
    \end{array}\right)
  \, .
\end{eqnarray}
In the CLBS we need for instance the fundamental operator insertions into the legs 1 and 2.
The components of the color matrix $\mbox{CF}$ are modified to
$(\mbox{CF}')_{11}=\mbox{Tr}[t^{b}t^{a}t^{c}t^{a}t^{b}t^{c}]$ and
$(\mbox{CF}')_{12}=\mbox{Tr}[t^{b}t^{a}t^{c}t^{b}t^{a}t^{c}]$.
Then the modified color matrix is obtained as 
\begin{eqnarray}
  \mbox{CF}' = \left( \begin{array}{rr}
      1/9 & 10/9  \\
      10/9 & 1/9  \\
    \end{array}\right).
\end{eqnarray}
One of the more complicated examples consists of the two color operator insertions into the 
process $g(a)g(b) \rightarrow t(1)\bar{t}(2)g(3)g(4)$. 
In MadGraph the normal $\mbox{SU}(3)$ color matrix for the process is a $24 \times 24$ matrix. 
The first 15 components in the first row read (we refrain from spelling out
the rest),
\begin{eqnarray}
    \label{eq:cf-ex}
  \mbox{CF} &=& \frac{1}{54}(512, 8, -64,80, 8, -10,-1, -64, -64, 8, -1, -10, -1, 62, -10 , \cdots ).
\end{eqnarray}
Upon insertions of two adjoint color operators for a gluon into the legs 3 and 4 the 
extended routines calculate the modified color matrix as
\begin{eqnarray}
    \label{eq:cfmod}
  \mbox{CF}' &=& \frac{1}{4}(8,0,8,16,0,-2,0,8,-1,-1,1,2,-8,-7,1, \cdots )\, ,
\end{eqnarray}
and, of course, the result in \Eq{eq:cfmod} agrees with independent checks.

\subsection{Spin correlations for gluon emitters}
\label{sec:gluo}
As we have seen above the dipoles for the splittings $g \to gg$ and $g \to q\bar q$ (which involve a gluon emitter)
introduce spin correlations (see e.g. \Eq{eq:dipggk}).
Since our numerical evaluation of the reduced Born amplitudes uses helicity amplitudes, 
we have to derive the components of the tensor for the spin correlations 
in the helicity formalism as well.
To that end, we choose the definitions in \cite{Xu:1986xb} which we here call the XZC
gluon polarization vector.

Let us illustrate the necessary steps (formulated in~\cite{Weinzierl:1998ki}) with the case 
of the massless dipole term $\Dipole(g_i,g_j,k)$ in \Eq{eq:dipggk},
\begin{eqnarray}
  \label{eq:dipggk-rev} 
  \Dipole(g_i,g_j,k) &=& -{1\over 2 p_i \cdot p_j}\ V_{g_ig_j,k}^{\mu\nu}\ \frac{1}{C_A} \left( A^{*}_{\mu}\ \T{k}\cdot\T{ij}\ A_{\nu} \right)
  \, ,
\end{eqnarray}
where $V_{g_ig_j,k}^{\mu\nu}$ is written schematically
as 
\begin{eqnarray}
\label{eq:vggk-rev}
  V_{g_ig_j,k}^{\mu\nu} &=& 16 \pi \al \ \mu^{2 \ep}\ C_A\ \left( -C_{1}g^{\mu \nu} +C_{2}L^{\mu}L^{\nu} \right) 
\, , 
\end{eqnarray}
with $L^{\mu}=z_{i}p_{i}^{\mu}-z_{j}p_{j}^{\mu}$ and some Lorentz scalars $C_{1,2}$  (e.g. given in \Eq{eq:vggk}).
Moreover, in \Eq{eq:dipggk-rev} we have expressed
the CLBS in \Eq{eq:clbs} through amplitudes $A_{\mu}$ and $A_{\nu}$, where the polarization
vector $\ep_{\mu}^{\lambda}(\tilde{p}_{ij})$ of the emitter gluon with
momentum $\tilde{p}_{ij}$ has been amputated. 
Then we can transform both the dipole splitting function $V^{\mu\nu}$ and the
amplitude $A_{\mu}$ to a helicity basis by inserting 
the polarization sum $\sum_{\lambda} \ep^{\lambda *}_{\mu}\ep^{\lambda}_{\nu}$ as
\begin{eqnarray}
A^{*}_\mu\ V^{\mu \nu}\ A_\nu &=& \sum_{\lambda',\lambda} A^{*}_{\lambda'}\ V^{\lambda' \lambda}\ A_{\lambda} 
\, , \\
\label{eq:splitcir}
V^{\lambda' \lambda} &=& \ep_{\mu}^{\lambda' *}\ V^{\mu \nu} \ep_{\nu}^{\lambda}
\, , 
\end{eqnarray}
and we obtain the dipole $\Dipole(g_i,g_j,k)$ in a $\pm$-helicity basis as
\begin{eqnarray}
\label{eq:dipggk-hel}
\Dipole(g_i,g_j,k) &=& -{1\over 2 p_i \cdot p_j} 
16 \pi \al \ \mu^{2 \ep}\ 
\\
& &
\times \
\left[
  \bigl( C_{1} + C_{2}|\mbox{E}_{+}|^{2}  \bigr) 
\left(A^{*}_+ \T{k}\cdot\T{ij} A_+ + A^{*}_- \T{k}\cdot\T{ij} A_-\right)
  + 2\ C_{2}\mbox{Re}\left(\mbox{E}_{+}^{*}\mbox{E}_{-} A^{*}_+ \T{k}\cdot\T{ij} A_- \right) \right] 
\, .
\nonumber
\end{eqnarray}
Here we have used gauge invariance, the on-shell condition for the gluon and orthogonality of the polarization vectors, 
which gives rise to the relations 
$\ep^{\lambda *}\cdot \tilde{p}_{ij} 
 = A \cdot \tilde{p}_{ij} 
 = L \cdot \tilde{p}_{ij} 
 = \tilde{p}_{ij}^2 = 0$. 
The quantity $\mbox{E}_{\pm}$ needed to express the dipoles in a helicity basis is defined as (see e.g. \cite{Weinzierl:1998ki})
\begin{eqnarray}
  \label{eq:Epm-def}
  \mbox{E}_{\pm} &=& \ep_{\pm} \cdot L 
  \, .   
\end{eqnarray}

As a subtlety, we would like to point out the following.
In the HELAS library~\cite{Murayama:1992gi,Hagiwara:1990dw} the gluon polarization vector is calculated 
by the subroutine {\tt VXXXXX} and taken to be in a helicity basis but the phase conventions 
are different from the ones in \cite{Xu:1986xb}.
Thus we have to relate these conventions to our choice~\cite{Xu:1986xb}.
Although it may be natural to use the same definition for the gluon polarization vectors as in HELAS for
the dipole splitting functions (e.g. $V_{g_ig_j,k}^{\mu\nu}$ in \Eq{eq:vggk-rev}) we have the freedom not to do so 
because of gauge invariance and the on-shell condition. 
In our implementation we have chosen the latter option 
with the advantage that the calculation of the dipole splitting functions can be completely separated from 
the part for evaluation of the CLBS. 
With the XZC definitions $\mbox{E}_{\pm}$ in \Eq{eq:Epm-def} is expressible in terms of spinor products 
which can be easily implemented in Fortran code.

The polarization vectors of HELAS~\cite{Murayama:1992gi,Hagiwara:1990dw} are
written in a helicity basis as
\begin{eqnarray}
  \label{eq:helas-def}
  \ep^{\pm}_{\mu [{\tiny \mbox{HELAS}}]}(k,q) &=& \frac{1}{\sqrt{2}}(\mp \ep_{\mu (1)}-i\ep_{\mu (2)})
  \, ,
\end{eqnarray}
with the explicit expression for the vectors in a linear basis, $\ep_{\mu (1,2)}$ 
given in~\cite{Murayama:1992gi,Hagiwara:1990dw}.
Likewise, for XZC~\cite{Xu:1986xb} we have 
\begin{eqnarray}
  \label{eq:xzc-def}
  \ep^{+}_{\mu [{\tiny \mbox{XZC}}]}(k,q)= \frac{ \bra q-| \gamma_{\mu}|k- \ket}{\sqrt{2} \bra q k \ket^{*}}
  \, , 
\end{eqnarray}
where the relation $\ep_{\mu [{\tiny \mbox{XZC}}]}^{+ *}=\ep_{\mu [{\tiny \mbox{XZC}}]}^{-}$ holds and 
$q$ is an arbitrary reference momentum.
The difference between the conventions \Eq{eq:helas-def} and \Eq{eq:xzc-def} amounts to a complex phase,  
\begin{eqnarray}
  \label{eq:xzc-helas-phase}
  \ep^{\pm *}_{[{\tiny \mbox{XZC}}]}(k,q) \cdot \ep^{\mp}_{[{\tiny \mbox{HELAS}}]}(k,q') &=& \mp e^{\pm i\phi(k)} 
  \, .
\end{eqnarray}
The phase difference contributes to only the second term in the square bracket in \Eq{eq:dipggk-hel} and it is 
then rewritten as
\begin{eqnarray}
  \label{eq:dipcir1s}
  + 2\ C_{2}\mbox{Re}\left(\mbox{E}_{+}^{*}\mbox{E}_{-} A^{*}_+ \T{k}\cdot\T{ij} A_- \right)_{[{\tiny \mbox{XZC}}]} &=&
  - 2\ C_{2}\mbox{Re}\left( \left(e^{+i\phi(k)}\ \mbox{E}_{+}\right)^{2} \left(A^{*}_+ \T{k}\cdot\T{ij} A_-\right)_{[{\tiny \mbox{HELAS}}]} \right)
  \, ,\qquad
\end{eqnarray}
where the CLBS is also taken in the helicity basis for the gluon emitter as in HELAS. 
The quantity $\mbox{E}_{+}=\ep^{+}_{[{\tiny \mbox{XZC}}]} \cdot L$ in \Eq{eq:dipcir1s} 
can be computed according to \Eq{eq:Epm-def} in the conventions of~\cite{Xu:1986xb} 
in terms of spinor products as (see e.g.~\cite{Weinzierl:1998ki}) 
\begin{eqnarray}
  \label{eq:E1}
  e^{+i\phi(k)} \mbox{E}_{+} &=& e^{+i \phi(\tilde{p}_{ij})}\ 
  \frac{z_{i}\bra p_{j} p_{i} \ket \ \bra \tilde{p}_{ij} p_{i} \ket^{*}}
  {\sqrt{2} \bra p_{j} \tilde{p}_{ij} \ket}
  \, ,
\end{eqnarray}
where we have set $k=\tilde{p}_{ij}$ and chosen $q=p_{j}$ for the reference momentum in \Eq{eq:xzc-def}. 
Explicit expressions for the vector $\mbox{E}_{+}$ in \Eq{eq:Epm-def}
for all required momentum configurations (massless and massive) are given in Appendix~\ref{sec:appB}. 
For the dipole terms including the massive partons, 
the result for $\mbox{E}_{+}$ in \Eq{eq:Epm-def} contains terms like 
$\bra p_{i}-|p_{k}|\tilde{p}_{ai}- \ket$ with on-shell momenta $p_{k}^2=m_{k}^{2}$. 
In order to express these terms through spinor products we have used flat momenta,
\begin{eqnarray}
p_{k}^{\flat} = p_{k} - \frac{m_{k}^{2}}{2p_{k} \cdot \tilde{p}_{ai1}} \tilde{p}_{ai},
\end{eqnarray}
which are massless because $(p_{k}^{\flat})^{2}=0$. 
Due to the equations of motion $\slash{\tilde{p}}_{ai} \ |\tilde{p}_{ai}- \ket =0$, we can rewrite 
the quantities under consideration, e.g.
\begin{eqnarray}
\bra p_{i}-| \ \slash{p}_{k} \ |\tilde{p}_{ai}- \ket &=&
\bra p_{i}-| \ \slash{p}_{k}^{\flat} \ |\tilde{p}_{ai}- \ket 
\, ,
\end{eqnarray}
so that they are accessible to standard spinor calculus, 
see also~\cite{Schwinn:2005pi} for the spinor helicity formalism including massive fermions.
Our treatment of the spin correlations is contained in the Mathematica sources 
where the expressions for $\mbox{E}_{+}$ (as given in Appendix~\ref{sec:appB}) 
are implemented. 
During the automated generation of all subtraction terms for numerical
evaluation the results are written to the output in the Fortran file, 
{\tt dipole.f}, see Fig.~\ref{fig:flowchart}.

\subsection{Checking the soft and collinear limits}
\label{sec:safe}
Let us discuss the checks for the AutoDipole package. 
First, there is the standard quality check on the automatic generation of the dipole terms 
because the result according to \Eq{eq:master} has to be finite when approaching the singular regions.
Second, since the Catani-Seymour formalism is well established, we can compare
the output of AutoDipole with results of independent implementations in the literature. 

The AutoDipole package performs automatically for all generated subtraction terms 
a numerical check of all IR limits.
For that purpose the subtracted squared matrix element for the process 
under consideration are  constructed and sampled over all limits to test the cancellation of 
the IR singularity.
For a given soft/collinear limit $\mbox{L}_i$, we pick up the set
$\mbox{S}(i)$ of the corresponding dipoles to test whether the quantity
\begin{eqnarray}
  \label{eq:check}
  \lim_{{\rm{L}}_i} \ \biggl[\  |\mbox{M}|^{2} - \sum_{j \subset {\rm{S}}(i)} \mbox{D}(j) \  
  \biggr],
\end{eqnarray}
is soft/collinear safe. 
The code lists all limits $\mbox{L}_i$, and the corresponding sets $\mbox{S}(i)$. 
A concrete example will be presented in Subsection~\ref{sec:short-example}. 
The leading soft/collinear singularity of the squared matrix element $|\mbox{M}|^{2}$ 
behaves in the soft limit as $1/k^{2}$ for a gluon of momentum $k$ and  
in the collinear one as $1/(2 p_{i} \cdot p_{j}) = 1/s_{ij}$ for two collinear momenta $p_{i}$ and $p_{j}$.
The set of the corresponding dipoles cancels this leading singularity pointwise as
\begin{eqnarray}
  \label{eq:mesqrd-minus-dip}
  |\mbox{M}|^{2} - \sum_{j \subset {\rm{S}}(i)} \mbox{D}(j) = 
  \frac{1}{x^2}(a_{0} + a_{1}x + a_{2}x^{2} + \cdots) - \frac{1}{x^2}a_{0}  
  \, ,
\end{eqnarray}
where $x=k(\sqrt{s_{ij}})$ is the respective Lorentz scalar depending on the
momenta in the soft (collinear) limit. 
The difference in \Eq{eq:mesqrd-minus-dip} is integrable over the real
emission phase space. 
The numerical accuracy, however, has to be controlled, because deep in the singular regions 
the individual contributions in \Eq{eq:mesqrd-minus-dip} become very large 
and with the limited numerical precision of floating point arithmetic the cancellation 
might be imperfect.
Fortunately, the stability of the numerical cancellation can be tested automatically.
Normalizing \Eq{eq:mesqrd-minus-dip} one can check the slope of the quantity,
\begin{eqnarray}
  \label{eq:mesqrd-minus-dip-norm}
  \frac{|\mbox{M}|^{2} - \sum \mbox{D} }{|\mbox{M}|^{2}} = \frac{1}{a_{0}}(a_{1}x + a_{2}x^{2}
  + \cdots )
  \, .
\end{eqnarray}
Based on \Eq{eq:mesqrd-minus-dip-norm} AutoDipole tests the cancellation and 
returns values for the fiducial regions for the Lorentz invariants $s_{ij}$.

We have extensively tested the soft and collinear finiteness of the generated subtraction terms 
for numerous scattering processes in $e^+e^-$, $ep$ and $pp$-collisions 
including massive quarks and weak gauge bosons. 
Among those are a large class of $2 \to 4$ scattering processes 
as needed in the NLO QCD correction to $2 \to 3$ reactions.
We have also tested scattering processes with $2 \to 5$ and $2 \to 6$ partons,
like e.g. 
\\[-3.0ex]
\begin{itemize}
\item[] $ u \bar{u} \rightarrow d \bar{d}ggg $ \\[-4.25ex]
\item[] $ gg  \rightarrow W^{+}\bar{u}dgg $ \\[-4.25ex]
\item[] $ gg  \rightarrow t\bar{t}ggg $ \\[-4.25ex]
\item[] $ gg  \rightarrow t\bar{t}b\bar{b}g $ \\[-4.25ex]
\item[] $ gg  \rightarrow t\bar{t}b\bar{b}gg $ \\[-3.0ex]
\end{itemize}
which are currently under investigation in view of phenomenological applications for the LHC.

In addition, we have been able to obtain also perfect agreement with published results~\cite{Dittmaier:2008uj,Dittmaier:2007th}.
For the NLO QCD corrections to $pp \rightarrow t\bar{t}+\mbox{1jet}$ production 
results for the real emission contributions have been presented at individual
phase space points in~\cite{Dittmaier:2008uj}.
We agree at least to 14 digits for the matrix elements squared and at least to 12 digits 
for the sum of the subtraction terms. 
All details of our comparison are shown in Table~\ref{tab:real1} in Appendix~\ref{sec:appC}. 
We have also checked against the results for the NLO QCD corrections to $pp \rightarrow W^{+}W^{-}+\mbox{1jet}$ \cite{Dittmaier:2007th}, 
in particular the real emissions in the channel $u{\bar u}\to W^{+}W^{-}gg$. 
Also here we have obtained very good agreement as documented in the last entry of Table~\ref{tab:real1}.
The AutoDipole package provides a shell script to reproduce some of the numbers in that Table~\ref{tab:real1} 
(see Section~\ref{sec:examples}).

Let us briefly comment on the integrated dipoles in \Eq{eq:master} originating from the $\mbox{\bf I}$-operators.
Here, a complete check is more involved because for a given process 
scheme dependence enters, i.e. whether the singularities of the NLO virtual contributions 
are factorized with respect to the $d$-dimensional or 4-dimensional Born amplitude.

Nevertheless, partial checks of the singularity structure are straightforward.
The matrix element squared with the sum of all $\mbox{\bf I}$-insertions is written schematically as 
\begin{eqnarray}
  \label{eq:Iterms-sketch}
  \bra 1,\ldots,n | \,\, \mbox{\bf I} \,\, | 1,\ldots,n \ket &=& 
  C_{-2} \frac{1}{\ep^2} + C_{-1} \frac{1}{\ep} + C_{0}
  \, ,
\end{eqnarray}
where $C_{-1}$ and $C_{-2}$ are process (and kinematics) dependent coefficients.
For the leading pole $ \sim 1/\ep^2$, the coefficient $C_{-2}$ obeys 
the following simple relation 
\begin{eqnarray}
\label{eq:Cdoublepole}
C_{-2} &=& 
\frac{\al}{2\pi} |\mbox{M}_{\tiny \mbox{Born}}|^2\, 
\left((n_{q}+n_{\bar q}) C_{F} + n_{g} C_{A} \right)
  \, , 
\end{eqnarray}
where $\mbox{M}_{\tiny \mbox{Born}}$ is the corresponding Born amplitude. 
The number of external massless (anti-)quarks is $n_{q}$ ($n_{\bar q}$) 
and $n_{g}$ the number of external gluons.
In QCD the standard $\mbox{SU}(N)$ color factors take the numerical values 
$C_{F}=4/3$ and $C_{A}=3$.

Apart from simple relations like \Eq{eq:Cdoublepole} the coefficients 
for color and spin averaged results for the {\bf I}-operator could again be
compared with the published literature for 
$pp \rightarrow t\bar{t}+\mbox{1jet}$ production at NLO in QCD~\cite{Dittmaier:2008uj}.
We have obtained agreement to at least 14 digits as shown in Table~\ref{tab:virtual1}.
The numbers in Table~\ref{tab:virtual1} can again be reproduced with a 
simple command explained in Section~\ref{sec:examples}.

%
%
\section{Running AutoDipole}
\label{sec:userguide}
%
%
This Section is meant to be a short manual to the AutoDipole package.
We explain the installation procedure, the use and discuss a few processes as examples.

\subsection{Install}
\label{sec:install}

The package, {\tt AutoDipole\_V1.2.3.tar}, is available for download from~\cite{AutoDipole:2009} 
(or else from the authors upon the request). 
In addition, one must also obtain the stand-alone version of MadGraph 
(MG\_ME\_SA\_V4.4.30) e.g. by download from~\cite{Madgraph:2009} 
and put it (in the form of {\tt .tar.gz} or {\tt .tar}) in the AutoDipole package directory.
Then execute the installation procedure by{\footnote{
\label{ft:mad-versions}
  The AutoDipole package (version 1.2.3 and earlier) is assumed to be
  used with MadGraph (stand-alone version MG\_ME\_SA\_V4.4.30). 
  Upgrades to newer versions of MadGraph require the user 
  to set the MadGraph version in the first line of {\tt install.sh} 
  by hand before installation.  
  We have tested that AutoDipole version 1.2.3 also works with 
  the MadGraph stand-alone version MG\_ME\_SA\_V4.4.39. 
  An automated test of the installation can be executed with the command {\tt check.sh} 
  as detailed in Section~\ref{sec:examples}.
}}
\begin{verbatim}
./install.sh
\end{verbatim}
and the directory structure as displayed in Figure~\ref{fig:fig1} emerges. 
Repetition of this command always allows the user to recover the initial
settings described in this Section.

\subsection{A short example}
\label{sec:short-example}

Let us next illustrate the use of AutoDipole. 
First of all, one has to specify all partonic real emission processes which appear 
at NLO in the observable under consideration.
To that end, as a concrete example let us choose a simple process, $u\bar{u} \rightarrow d\bar{d}g$ 
which contributes to hadronic di-jet production, $pp \rightarrow
\mbox{2~jets}$ at NLO. 
This example exhibits all features of the Catani-Seymour formalism, i.e. it needs dipoles and {\bf I}-operators as well
as the {\bf P}- and {\bf K}-terms.
We start with the Mathematica part of AutoDipole. 
The package can simply be included through the driver file as
\begin{verbatim}
<<driver_user.m
\end{verbatim}
Next, we can run the package for the process $u\bar{u} \rightarrow d\bar{d}g$ 
with the command
\begin{verbatim}
GenerateAll[{u,ubar},{d,dbar,g}]
\end{verbatim}

Upon running the AutoDipole package successfully, all generated Fortran files 
are stored in a (newly created) directory under {\tt process}, 
for the example at hand {\tt ./process/Proc\_uux\_ddxg}. 
AutoDipole returns the message:
\begin{verbatim}
******************************************************'
Run has been succeeded.
In order to run the generated code, please enter
in the newly created directory under ./process
make
./runD
******************************************************'
\end{verbatim}

The real matrix element squared $|\mbox{M}|^{2}$ and the sum of all dipole terms $\sum_{i} \mbox{D}(i)$
are evaluated at 10 phase space points with this command.
Also the check of cancellations in all infrared limits for the
subtracted matrix element can performed. 
The respective command is:
\begin{verbatim}
make checkIR
./checkIR
\end{verbatim}

Likewise, the integrated dipoles ({\bf I}-operators) as well as the
contribution of the {\bf P}- and {\bf K}-terms at 10 points in phase space 
can be evaluated in the subdirectories {\tt Proc\_uux\_ddxg/Virtual} and {\tt Proc\_uux\_ddxg/PK} 
with the commands:
\begin{verbatim}
make runI
./runI
\end{verbatim}
\begin{verbatim}
make runPK
./runPK.
\end{verbatim}
The present example $u\bar{u} \rightarrow d\bar{d}g$ runs with the default settings of AutoDipole. 
For further illustration the package also comes along with a prepared list of examples so that 
explicit numbers can be obtained by executing the shell scripts explained in the next subsection.

\begin{table}[htbp]
  \centering
\begin{tabular}[c]{|l|l|l|} 
\hline
{\bf file (subroutine)}
  & {\bf input}
  & {\bf output}
\\
\hline
&&\\
$[$Dipole terms$]$
&&\\
{\tt matrix.f} (smatrix)               &$\{p_i\}$                        & $|\mbox{M}|^2$ for real emission \\
{\tt dipole.f} (dipole)                &$\{p_i\}$                        & $\mbox{D}(i)$ \\
{\tt reducedm.f} (reducedm)            &$\{p_i\}$                        & reduced $\{{\tilde p}_{i}\}$ \\ 
{\tt cmatrix.f} (cmatrix\#)            &reduced $\{{\tilde p}_{i}\}$     & CLBS (like helicity) \\
{\tt cmatrix.f} (cmatrix\#dh)          &reduced $\{{\tilde p}_{i}\}$     & CLBS (unlike helicity) \\
{\tt check.f} (program checkdipole)    &$\{p_i\}$ in {\tt inputm.h}      & check of $\mbox{D}(i)$, $|\mbox{M}|^2$ \\
&&\\
$[${\bf I}-terms$]$
&&\\
{\tt matrixLO1.f} (smatrix)            &$\{p_i\}$                        & $|\mbox{M}_{\tiny \mbox{Born}}|^2$ \\
{\tt Iterm.f} (Iterm)                  &$\{p_i\}$                        & $C_i$ of {\bf I}-terms  \\
{\tt cmatrix.f} (cmatrix\#)            &$\{p_i\}$                        & CLBS \\
{\tt checkVirt.f} (program checkVirt)  &$\{p_i\}$ in {\tt inputmLO.h}    & check of $C_i$ of {\bf I}-terms  \\
&&\\
$[${\bf P}- and {\bf K}-terms$]$
&&\\
{\tt matrixLO1.f} (smatrix)            &$\{p_i\}$                        & $|\mbox{M}_{\tiny \mbox{Born}}|^2$ \\
{\tt PKterm.f} (PKterm)                &$\{p_i\}$                        & $\mbox{Pt}(i)$, $\mbox{K}(i)$ ({\bf P}- and {\bf K}-terms)\\
{\tt cmatrix.f} (cmatrix\#)            &$\{p_i\}$                        & CLBS \\
{\tt checkPK.f} (program checkPK)      &$\{p_i\}$ in inputmLO.h          & check of $\mbox{Pt}(i)$ and $\mbox{K}(i)$ \\
&&\\
$[$Check of soft/collinear limits$]$
&&\\
{\tt collinear.f}                      & $\{p_i\}$                        & collinear limits of $\{p_i\}$ \\
{\tt checkIR.f} (program checkIR)      & collinear limit of $\{p_i\}$     & $(|\mbox{M}|^{2} - \sum \mbox{D})/|\mbox{M}|^{2}$ \\
&&\\
\hline
{\bf common input files}
&\multicolumn{2}{l|}{} \\
\hline
&\multicolumn{2}{l|}{} \\
$[$Process dependent$]$
&\multicolumn{2}{l|}{} \\
{\tt allcolormat\#.inc}                 &\multicolumn{2}{l|}{extended color matrices (used in cmatrix\#)} \\
{\tt emitspecinfo.inc}                  &\multicolumn{2}{l|}{information of emitter for dipoles (used in cmatrix\#)} \\
{\tt param\_card.dat}                   &\multicolumn{2}{l|}{parameters for CLBS} \\
{\tt nexternal.inc}                     &\multicolumn{2}{l|}{field numbers of real emission process for dipoles and} \\ 
                                        &\multicolumn{2}{l|}{of Born process for
                                           {\bf I}-, {\bf P}- and {\bf K}-terms} \\
{\tt nexternal2.inc}                    &\multicolumn{2}{l|}{field number of reduced Born process} \\
{\tt inputm.h}                          &\multicolumn{2}{l|}{phase space points for dipole checks} \\
{\tt inputmLO.h}                        &\multicolumn{2}{l|}{phase space points for {\bf I}-, {\bf P}-, and {\bf K}-term checks} \\
&\multicolumn{2}{l|}{} \\
$[$Process independent$]$
&\multicolumn{2}{l|}{} \\
{\tt coupl.inc}                         &\multicolumn{2}{l|}{common block of parameters} \\
{\tt couplings.f}                       &\multicolumn{2}{l|}{reads parameters in {\tt param\_card.dat} and assigns} \\
                                        &\multicolumn{2}{l|}{them in {\tt coupl.inc}} \\
&\multicolumn{2}{l|}{} \\
\hline
\end{tabular}
\caption{\small
\label{tab:file-content}
The file content of the AutoDipole package. In names of subroutine and files 
(e.g. cmatrix\#) the symbol \# denotes an integer number, typically \#=1.
}
\end{table}
If the user wants to include the generated Fortran code with the subtraction terms 
into his/her own project for further numerical evaluations, he/she needs (i.e. has to
copy) a number of files for that purpose. The complete list of generated files
is given in Table~\ref{tab:file-content}. 
Note that most of the functions are only needed internally and are not meant
to be called by the user directly. 
We assume here that all input momenta $\{p_i\}$ are generated from phase space
routines supplied by the user and that the HELAS library {\tt libdhelas3.a} is linked.
\\

{\bf Dipole terms :} directory {\tt process/Proc\_uux\_ddxg} \\
From the file {\tt check.f} the call of the relevant subroutines for the
evaluation of the dipole terms is obvious. 
The user needs to copy the following files:
\begin{verbatim}
matrix.f                      (generated by AutoDipole)
dipole.f                                 |
reducedm.f                               |
cmatrix.f                                |
allcolormat\#.inc                        |
emitspecinfo.inc                         |
nexternal.inc                            |
nexternal2.inc                (generated by AutoDipole)

coupl.inc                     (process independent file)
couplings.f                   (process independent file)

param_card.dat                (parameter input file)
\end{verbatim}
To use the generated code, one first has to initialize MadGraph.
The parameters are determined in {\tt param\_card.dat} and are assigned 
to the definitions in {\tt coupl.inc} 
with the help of the subroutine {\tt setpara} in {\tt couplings.f} as
\begin{verbatim}
include 'coupl.inc'
call setpara('param_card.dat',.true.)
\end{verbatim}
to be used subsequently in the subroutines {\tt cmatrix\#}, {\tt cmatrix\#dh}, and {\tt smatrix} 
through the common blocks. 
The symbol \# here and above denotes an integer number (e.g. \#=1).
For the dipoles only the strong coupling constant $\al$ as well as the top and bottom masses 
are needed in the evaluation. 
These three parameters appear as
\begin{verbatim}
AL=g**2*(4.d0*pi)**(-1)
mt=tmass
mb=bmass
\end{verbatim}
and are passed to the subroutine {\tt dipole} through the common blocks,
\begin{verbatim}
common /usedalpha/ AL
common /MASS/ mt,mb
\end{verbatim}
Next, the evaluation of the squared matrix element for the real emission process $|\mbox{M}|^2$ 
proceeds in exactly the same way as in MadGraph with the call
\begin{verbatim}
call smatrix(p,msq)
\end{verbatim}
Finally, the dipole terms are evaluated by calling the subroutine
\begin{verbatim}
call dipole(p,dip,SumD)
\end{verbatim}
which returns an array with the values for each dipole term ({\tt dip}) 
and the sum of the all dipoles ({\tt SumD}) as output.
For our example process $u\bar{u} \rightarrow d\bar{d}g$ we have 15 individual dipoles.
The reduced kinematics of each dipole (as determined by the subroutine {\tt reducedm}) 
can be accessed by the common block 
\begin{verbatim}
double precision ptil(0:3,1:n1,n2)
common /OUTPUTM/ ptil
\end{verbatim}
where {\tt n1}=4 and {\tt n2}=15 in the definition of the momenta for the example at hand.
\\

{\bf {\bf I}-terms :} directory {\tt process/Proc\_uux\_ddxg/Virtual} \\
Our program {\tt checkVirt.f} 
for checks of the {\bf I}-terms again provides an example 
for the use of the subroutines to evaluate the {\bf I}-terms.
In addition to the eight files from {\tt cmatrix.f} to {\tt param\_card.dat}
listed above for the evaluation of the dipoles, 
the user has to copy one file:
\begin{verbatim}
Iterm.f
\end{verbatim}
Then, the initialization phase is exactly the same as before and 
for the evaluation of the {\bf I}-terms, one has to call the subroutine 
\begin{verbatim}
call Iterm(p,coef,SumI)
\end{verbatim}
which provides the value of the coefficients (i.e. the array {\tt coef}) as well as the 
sum ({\tt SumI}) of all {\bf I}-terms as output. 
If the user also wants to compute the LO matrix element $|\mbox{M}_{\tiny \mbox{Born}}|^2$ 
he needs the file {\tt matrixLO1.f}. 
The execution of the subroutine with the command 
\begin{verbatim}
call smatrix(p,msqLO)
\end{verbatim}
is performed again exactly as in MadGraph.
\\

{\bf {\bf P}- and {\bf K}-terms :} directory {\tt process/Proc\_uux\_ddxg/PK}  \\
The user needs the following file to evaluate the {\bf P}- and {\bf K}-terms:
\begin{verbatim}
PKterm.f
matrixLO1.f
\end{verbatim}
as well as the eight files from {\tt cmatrix.f} to {\tt param\_card.dat} 
already discussed for the dipoles.
The initialization phase is unchanged but the user has to provide the parton
momentum fraction $x$ from the mass factorization as additional input.
The {\bf P}- and {\bf K}-terms for the first leg are computed by the call
\begin{verbatim}
call PKterm1(p,x,SumP,SumK)
\end{verbatim}
which returns separately the sum of all {\bf P}- and {\bf K}-terms ({\tt SumP} and {\tt SumK}).
For the second parton in initial state (as in hadron-hadron collisions) 
the corresponding {\bf P}- and {\bf K}-terms are evaluated by calling {\tt PKterm2}.

\subsection{General usage}
\label{sec:use}
For a specific application of AutoDipole, the user needs to perform the following three steps:
\begin{verbatim}
1. Setup of parameters : parameter.m
2. Run of package      : GenerateAll[{initial},{final}]
3. Run and checks of the generated code at : /process/Proc_xx_xx/ .
\end{verbatim}
Let us explain each of these steps in more detail.

\vspace{5mm}
{{\bf 1. Setup of parameters}}\\
Generally, the user has to supply all process parameters needed 
for the dipole subtraction procedure by editing the file,
\begin{verbatim}
parameter.m
\end{verbatim}
where (following the MadGraph conventions) 
all values with the dimension of a mass are in units of [GeV].
Let us explain all variables in the file together with the chosen default values.

\bigskip

\noindent
{\bf Parameters for dipole terms}\\[-4ex]
\begin{itemize}
\item[] {\tt ep}=0 : \\
  The parameter of dimensional regularization of space-time, $D=4-2\epsilon$.
  The default choice eliminates higher orders (positive powers) of $\epsilon$
  in the dipole terms.

\item[] {\tt kap}=2/3 :\\ 
  $\kappa$ is free parameter in some expressions for massive dipoles~\cite{Catani:2002hc}.
  The value $\kappa=0$ leads to the simplest results for the dipole terms,
  while $\kappa=2/3$ produces the simplest expressions of the ${\mbox{\bf I}}$-terms.

\item[] {\tt skipdipole}=\{ \} : \\
  This set specifies the kind of dipoles to be skipped during the creation. 
  For example, the set {\tt skipdipole}=\{2t\} 
  omits the creation of the Dipole 2-(5) with the $t-\bar{t}$ splitting
  (see Table~\ref{tab:creation-order-dip} in Appendix~\ref{sec:appA}).

\item[] {\tt mur}=174.30 : \\
  The value of the renormalization scale. 
  The default value is the same as the default value of the top-quark mass in MadGraph.

\item[] {\tt acccut}=10$^{\wedge}$(-3) : \\ 
  This value is used for the checking the quality of the infrared cancellations 
  in the generated code upon scanning over all possible collinear configurations.
\end{itemize}

\noindent
{\bf Parameters for ${\mbox{\bf I}}$-terms}\\[-4ex]
\begin{itemize}
\item[] {\tt mFlist}=\{ \} : \\
  This set lists all heavy quarks which possibly contribute to the ${\mbox{\bf I}}$-terms 
  with a gluon emitter in final state, e.g. as {\tt mFlist}=\{t,b\}.

\item[] {\tt lightflavors}=1 : \\
  This is the number of the light flavors which contributes to the ${\mbox{\bf I}}$-terms
  with a gluon emitter in initial and final states.

\item[] {\tt replistVirtual}=\{ \} : \\
  This set enables user defined symbolic replacements to simplify the virtual
  contribution, 
  e.g. of the type, 
  {\tt replistVirtual}=\{Gamma[1-eps]-> (4*Pi)$^{\wedge}$(eps)*somesymbol$^{\wedge}$(-1)\}
  which results in the replacement, $\frac{(4\pi)^{\ep}}{\Gamma(1-\ep)}=$somesymbol.

\item[] {\tt sijexpansion}=0 : \\
  This parameter determines whether the typical factors of dimensional
  regularization like $\left( {4 \pi \mu^2 \over 2 p_i \cd p_j} \right)^\ep$
  in the ${\mbox{\bf I}}$-terms are expanded~(0) or not~(1).
\end{itemize}

\noindent
{\bf Parameters for ${\mbox{\bf P}}$- and ${\mbox{\bf K}}$-terms}\\[-4ex]
\begin{itemize}
\item[] {\tt KFSff}=0, {\tt KFSgg}=0, {\tt KFSfg}=0, and {\tt KFSgf}=0 : \\
  Theses parameters specify the factorization scheme. 
  In the $\overline{\mbox{MS}}$ scheme (default choice) they are all vanishing.

\item[] {\tt lightquarksPK3}=\{u\} : \\
  This set specifies all light flavors to be added to the initial
  gluon of the reduced Born process 
  in the ${\mbox{\bf P}}$- and ${\mbox{\bf K}}$-terms (see group (6) 
  in Table~\ref{tab:creation-order-pandk} in Appendix~\ref{sec:appA}).

\item[] {\tt muFfort}=174.30 : \\
  This is the value of the factorization scale $\mu_F$, 
  which has the same default value as the renormalization scale $\mu_R$.
\end{itemize}

The other parameters which are used by MadGraph are determined in the file \\
{\tt patch/Models/sm/param\_card.dat}. \\
The default values for the most important ones are the following (in units of [GeV] for masses),
\\[-3.0ex]
\begin{itemize}
\item[] $\al$=0.1180\\[-4.25ex]
\item[] $\alpha_{em}$=0.007547169811320755\\[-4.25ex]
\item[] Top mass=174.3\\[-4.25ex]
\item[] Bottom mass=4.7\\[-4.25ex]
\item[] W mass=80.419\\[-4.25ex]
\item[] Z mass=91.188\\[-4.25ex]
\item[] Higgs mass=120.0\\[-3.0ex]
\end{itemize}
These parameters are used consistently throughout the run of the package.
Any other changes of SM parameters like e.g. $G_F$, the CKM-parameters, the
decay widths of heavy fields, and so on, are done in the same MadGraph file, {\tt param\_card.dat}.
Similarly, also control over all interactions in the SM
proceeds entirely through MadGraph, i.e. through the file, \\
{\tt patch/Models/sm/interactions.dat} \\
in the exactly same way as in the normal use of MadGraph.
All MadGraph parameters are explained in~\cite{Madgraph:2009}.

\vspace{5mm}
{{\bf 2. Run of package}}\\
For general QCD processes (including leptons and SM bosons) 
we run AutoDipole by executing the Mathematica command:
\begin{verbatim}
GenerateAll[{initial},{final}]
\end{verbatim}
where the sets labeled {\tt initial} and {\tt final} contain all external
particles of the respective scattering process.
The following fields are available in the present version of the package:
\begin{enumerate}
\item[] Fermions: \\  ($u, d, b, t, e^{-}, \mu^{-} ,\bar{u}, \bar{d}, \bar{b}, \bar{t}, e^{+}, \mu^{+}$) = \\
  ({\tt u, d, b, t, e, muon, ubar, dbar, bbar, tbar, ebar, muonbar}) \, , 
\item[] Bosons: \\ 
  ($gluon, photon, W^{+}, W^{-}, Z, Higgs$) = ({\tt g, gamma, Wp, Wm, Z, h})\, ,
\end{enumerate}
where the right hand side denotes the input format of AutoDipole.
The available quark fields include the first family ($u$ and $d$-quarks) 
which serves as a template for light quarks and, 
respectively, the third family ($b$ and $t$-quarks) as one for heavy quarks.
In practice this suffices as far as the subtraction formalism is concerned, 
because generally QCD scattering processes at NLO will be insensitive 
to the individual flavor of the quarks, 
and the sum over all light quark flavors can be trivially taken afterwards.
If one intends to study particular channels, e.g. in $W^\pm$ production, 
flavor separation is required, though.

As a result of the run, AutoDipole produces the following information 
about the created dipoles and the ${\mbox{\bf I}}$-, ${\mbox{\bf P}}$- and 
${\mbox{\bf K}}$-terms:
\begin{itemize}
\item[--] 
  All analytical expressions for the dipole terms are written in the
  order given in  Table~\ref{tab:creation-order-dip} along with the reduced Born process 
  (emitter,spectator) and the reduced kinematics for each dipole term.
  Also, the number of created dipole terms is shown group by group, namely Dipole 1, 2, 3, and 4
  and their subgroups (see Figure~\ref{fig:layout-dip}). 
  The number of massive dipole terms is given separately.

\item[--] 
  All soft and collinear limits together with the corresponding dipole terms
  are returned.

  For example, for the process $g(1)g(2) \rightarrow t(3)\bar{t}(4)g(5)g(6)$
  (see next Subsection)
  the fifth collinear limit is denoted $\mbox{Lc}(5)=(5,6) : \mbox{Sc}(5)=\{17, 18, 19, 20\}$.
  This means that the collinear divergence of the pair (5(g),6(g)) is canceled by 
  the set of the dipoles, $\mbox{Sc}(5)$  
  and the set consists of the dipoles number 17,18,19, and 20.
  In the same way, the soft limit and the corresponding dipoles are displayed. 
  E.g. for the soft limit of gluon 5(g), 
  $\mbox{Ls}(1)=(5) : \mbox{Ss}(1)=\{\mbox{Sc}(1), \mbox{Sc}(2), \mbox{Sc}(5), \mbox{Sc}(6), \mbox{Sc}(7)\}$, 
  which means that the soft divergence of 5(g) is canceled by the set of the dipoles, $\mbox{Ss}(1)$.
  This set in turn consists of the sets of collinear limits $\mbox{Sc}(1)$, $\mbox{Sc}(2)$, and so on.
         
\item[--] 
  The analytical expressions of all ${\mbox{\bf I}}$-terms 
  are shown in the order as given in Table~\ref{tab:creation-order-I} 
  prior to the expansion in $\ep$ 
  as well as the individual coefficients of the poles $C_{-2},\, C_{-1}$, and
  $C_{0}$ (see \Eq{eq:Iterms-sketch}) 
  after the $\ep$ expansion. 
  Also the total number of all created ${\mbox{\bf I}}$-terms is displayed 
  by group by group (see Table~\ref{tab:creation-order-I}).

\item[--] 
  The analytical expressions of all created ${\mbox{\bf P}}$- and ${\mbox{\bf K}}$-terms are shown 
  in the order given in Table~\ref{tab:creation-order-pandk} together with the
  total number of ${\mbox{\bf P}}$- and ${\mbox{\bf K}}$-terms.

\item[--] 
  Finally, AutoDipole outputs a summary with the numbers of all created dipoles, ${\mbox{\bf I}}$-, ${\mbox{\bf P}}$- 
  and ${\mbox{\bf K}}$-terms and the reduced Born processes.
\end{itemize}
This information, being rather detailed, helps with the identification of individual terms according to the order of creation.
The Mathematica part of AutoDipole provides analytical and symbolic expressions, which are easily accessible, 
if needed. These features are primarily for an experienced user.

\bigskip

Upon running the AutoDipole package it occurs (on purely algorithmic grounds) 
in some cases though, that a particular reduced Born cross section does not exist. 
For example, the process $e^{-}e^{+} \rightarrow u\bar{u}g$  does give rise through the dipole 2-(5)-($u,\bar{u}$) 
to the Born process $e^{-}e^{+} \rightarrow gg$, which does not exist at tree level.
In such cases, an error message is displayed:
\begin{verbatim}
Reduced Born amplitude B2u does not exist.
Dipole 2u must be switched off.
Please set in the file ./parameter.m 
skipdipole={ 2u }
\end{verbatim}
Following the message the user has to modify the command {\tt skipdipole} and repeat the Mathematica run. 
In this way only the relevant partial subsets of dipole terms is extracted and
the offending Born processes are skipped (thanks to {\tt skipdipole}) before calling the MadGraph part.

\vspace{5mm}
{{\bf 3. Run and checks of the generated code}}\\
Finally, as briefly explained above, the user can run the generated Fortran code for the regularized 
real emission process as well as the ${\mbox{\bf I}}$-, ${\mbox{\bf P}}$- and ${\mbox{\bf K}}$-terms.
The checks proceed as follows:

\bigskip

\noindent
{\bf Dipole terms}\\[-4ex]
\begin{itemize}
\item[] 
As mentioned above, the evaluation of the real matrix element squared $|\mbox{M}|^{2}$ and the sum of all dipole terms $\sum_{i} \mbox{D}(i)$
at 10 randomly chosen phase space points as well as the check of cancellations in all infrared limits for the
subtracted matrix element are automatically executed:
\begin{verbatim}
make 
./runD

make checkIR
./checkIR
\end{verbatim}
The latter command performs a scan over all possible collinear configurations.
For the expression 
\begin{equation}
  \label{eq:irsafe-check}
  {|\mbox{M}|^{2} - \sum \mbox{D} \over |\mbox{M}|^{2}}
  \, ,
\end{equation}
cf. \Eq{eq:mesqrd-minus-dip-norm}, 
it is checked whether the cancellation works at least to the accuracy defined by the value of the parameter {\tt acccut} 
in the file {\tt parameter.m}.
If the absolute value of the normalized subtracted squared matrix element in \Eq{eq:mesqrd-minus-dip-norm} 
drops below the pre-defined value of {\tt acccut} sufficiently deep in the collinear limit, 
the message, 
\begin{verbatim}
Infrared safety of all collinear limits is confirmed for S_ij/S > ...
\end{verbatim}
is printed. 
\end{itemize}

\noindent
{\bf ${\mbox{\bf I}}$-terms}\\[-4ex]
\begin{itemize}
\item[]
The Fortran code to evaluate all ${\mbox{\bf I}}$-terms is collected in the subdirectory {\tt ./Virtual}.
There, the coefficients of the double and single poles in the $\epsilon$
expansion $C_{-2}, \ C_{-1}, \ C_{0}$ in \Eq{eq:Iterms-sketch} can be evaluated 
at 10 randomly chosen phase space points by:
\begin{verbatim}
make runI
./runI
\end{verbatim}
This includes in particular the calculation of the double poles according to  
\begin{eqnarray}
\frac{C_{-2}}{(\al/2\pi) |M_{\tiny \mbox{Born}}|^2}\, ,
\end{eqnarray}
and the user can easily confirm that \Eq{eq:Cdoublepole} holds.

\end{itemize}

\noindent
{\bf ${\mbox{\bf P}}$- and ${\mbox{\bf K}}$-terms}\\[-4ex]
\begin{itemize}
\item[] 
Likewise, the mass factorization terms originating from the ${\mbox{\bf P}}$ and ${\mbox{\bf K}}$-terms is stored in {\tt ./PK}
and also evaluated at 10 randomly chosen phase space points by the command:
\begin{verbatim}
make runPK
./runPK
\end{verbatim}

\end{itemize}

This concludes our brief description of the user specific features of the AutoDipole package.

\subsection{Examples}
\label{sec:examples}

The package includes a command to demonstrate the run for several example processes. 
This helps to explain the package and at the same time confirms that the installation 
was done successfully and that the package works properly.
These examples include the computation of some of the numbers given 
in Tables~\ref{tab:real1} and \ref{tab:virtual1} in Appendix~\ref{sec:appC}. 
The command 
\begin{verbatim}
./lib/sh/check.sh
\end{verbatim}
generates five processes in different kinematics, 
\\[-3.0ex]
\begin{itemize}
\item[1.] $ e^{-} e^{+} \rightarrow u\bar{u}g $ \\[-4.25ex]
\item[2.] $ e^{-} u \rightarrow e^{-}ug $ \\[-4.25ex]
\item[3.] $ gg  \rightarrow t\bar{t}gg $ \\[-4.25ex]
\item[4.] $ ug  \rightarrow t\bar{t}ug $ \\[-4.25ex]
\item[5.] $ u\bar{u} \rightarrow W^{+}W^{-}gg $ \\[-3.0ex]
\end{itemize}
provides suitable settings for the parameters of each process and runs the package.
For illustration, we only explain in the following those settings which are
changed with respect to the default ones.

The processes from 1 to 4 use a value of $\al$=0.1075205492734706 
from~\cite{Dittmaier:2008uj} for the strong coupling constant.
For the processes 1 and 2 the contribution of the $Z$-boson exchange is switched off 
in the file {\tt interactions.dat} for simplicity. 
Moreover, in {\tt parameter.m} the splitting 2u in process 1 (and 3u in 2) 
is skipped by the command {\tt skipdipole = \{2u\}} (and {\tt skipdipole =\{3u\}}
in 2, respectively) 
to exclude an inexistent reduced Born process. 

For the processes 3 and 4 the QED interaction is switched off 
in the file {\tt interactions.dat} in order to focus on the QCD corrections only.
Furthermore, a value of 174 [GeV] for the top mass is used following~\cite{Dittmaier:2008uj} 
and the decay width of the top-quark must be switched off in the file {\tt param\_card.dat}.
For the comparison with~\cite{Dittmaier:2008uj} the $t-\bar{t}$ splitting is
skipped in {\tt parameter.m}, i.e. {\tt skipdipole =\{2t\}}, 
and the renormalization and factorization scales are set equal to the top mass value.
For the ${\mbox{\bf I}}$-terms five light flavors are taken into account, 
{\tt lightflavors = 5}.

Finally, process 5 uses the following settings~\cite{Dittmaier:2007th} in
the file {\tt param\_card.dat}: $W$-mass=80.425, 
$\al$= 0.1202629003906369, $\alpha_{em}$=0.007543595708669335 
and Fermi's constant $G_{F}=1.16637 \times 10^{-5}$, 
and the decay width of $W$-boson is switched off.
The results of processes 3, 4, and 5 should reproduce the corresponding numbers 
in Tables~\ref{tab:real1} and \ref{tab:virtual1}.

The test run of all examples is finally checked by comparing all computed values 
of the dipoles, {\bf I}-terms, and the averaged matrix element squared $|\mbox{M}|^2$ 
for the processes 1 to 5 the with ones in the data files in {\tt./lib/check/}.
At the end of the run, e.g. for example 3, the following message appears:
\begin{verbatim}
 3.gg_ttxgg
   |M|^2(Real) :Confirmed
    Dipole     :Confirmed
   |M|^2(LO)   :Confirmed
    I-term     :Confirmed
\end{verbatim}
This confirms the successful installation and proper functioning of the package. 
The generated code for all examples is stored in the directory {\tt ./process} 
as in a usual run and the directory of each process includes in the file {\tt res\_std} 
the result of the comparison with the data files from {\tt./lib/check/}.
All parameters used for each process are accessible in {\tt param\_card.dat}, 
so the user can also run the respective files individually with usual commands of the previous subsection. 

%
%
\section{Conclusion}
\label{sec:conclusions}
%
%
We have presented the package AutoDipole which provides an implementation of the Catani-Seymour dipole
formalism~\cite{Catani:1996vz,Catani:2002hc} to compute subtracted matrix elements squared in an automatic way. 
The package (partial aspects of which have been discussed before~\cite{Hasegawa:2008ae,Hasegawa:2009zz})
consists of essentially three ingredients: 
the automatic generation of all dipole terms via Mathematica routines as well
as the calculation of the color-correlated matrix elements 
and the evaluation of different helicity amplitudes with the help of 
MadGraph~\cite{Stelzer:1994ta,Maltoni:2002qb,Alwall:2007st} (i.e. a patched stand-alone version).

We have presented a complete discussion of the set-up of the program and the details of our
implementation as well as the numerical evaluation of the squared matrix elements.
Particular emphasis has been put on explaining the use of AutoDipole which
has been illustrated with sufficiently non-trivial examples.
The finiteness in the soft and collinear limits has been demonstrated for a
large number of processes in $e^+e^-$, $ep$ and $pp$-scattering.
Also, checks for the complete set of dipoles have been performed 
by comparing to various processes in the recent literature.

In comparison with other available software to generate subtraction terms 
for the real emission contributions AutoDipole has a number of advantages.
First of all, the algorithm is implemented in a way so that analytic expressions are generated
which are accessible to symbolic manipulation. 
Then, the chosen order for the dipole generation produces flat Fortran code 
and allows to handle scattering processes with $2 \to 5$ and $2 \to 6$ partons 
of relevance for LHC phenomenology including the computation of the CLBS.
Finally, the structure of the package is highly modular and flexible for 
further extensions.

Future improvements of the AutoDipole package will address restricting the dipoles 
to small regions of phase space, as realized e.g. through $\alpha$-parameters~\cite{Nagy:1998bb}. 
This is advantageous when performing the Monte Carlo integration over the 
phase space of the real emission process. 
The implementation of such cuts affects also the integrated dipoles and the
case of massless partons is available in the literature 
whereas for the case of massive partons most cases can be found in~\cite{Campbell:2004ch,Campbell:2005bb}.
The remaining integrals for exhaustive coverage can be done e.g. with the methods of~\cite{Bolzoni:2009ye}.
Next, the current version of AutoDipole does not cover processes with
identified hadrons in the final state where one needs fragmentation.
Here, the so-called ${\mbox{\bf H}}$-terms have to be taken into account, 
which will be implemented in a later version of AutoDipole. 
Finally, the structure of our code is particularly suited 
for extensions to physics beyond the Standard Model (BSM). 
The main modification needed is the so-called model file of MadGraph, 
which is currently already available for a large class of BSM scenarios 
including e.g. the minimal supersymmetric extension of Standard Model 
for which the necessary modifications of the dipole formalism are known~\cite{Catani:2002hc}.
All the extensions mentioned above can be easily realized within the existing
structure of our code.

\smallskip

The AutoDipole package is available for download from~\cite{AutoDipole:2009} or from the authors upon request.
The MadGraph package (stand-alone version) which includes the HELAS library may be obtained from~\cite{Madgraph:2009}.

\subsection*{Acknowledgments}
We acknowledge useful discussions with S.~Badger, T.~Diakonidis, R.~Frederix, F.~Maltoni, T.~Riemann, and B.~Tausk.
This work is supported in part by the Deutsche Forschungsgemeinschaft in Sonderforschungs\-be\-reich/Transregio~9,
the European Community in Marie-Curie Research Training Network MRTN-CT-2006-035505 {\it ``HEPTOOLS''}, 
and the Helmholtz Gemeinschaft under contract VH-NG-105 and contract VH-HA-101
(Alliance {\it ``Physics at the Terascale''}).

{\footnotesize


}

\newpage
\appendix
%
%
\section*{Appendix}
%
%

\section{Creation order of subtraction terms}
\label{sec:appA}
\renewcommand{\theequation}{\ref{sec:appA}.\arabic{equation}}
\setcounter{equation}{0}

\begin{table}[h!]
  \centering
\begin{eqnarray}
\begin{array}{|l|l|l|l|l|} \hline
  \ \ \ \ \mbox{Emitter}    
& \mbox{Spectator} 
& \mbox{Dipole} 
& \ \  \mbox{Condition} 
& \ \mbox{Eq.} 
\\ \hline
\mbox{Dipole 1}    &  &  & & 
\\ \hline
   (1) \ (i,j)=(f,g)  & 1. \ k & \mbox{D}_{ij,k}   & m_{i}=m_{k}=0
   & (5.7) \mbox{\ in \ \cite{Catani:1996vz}} \\ \hline
                      &                    & \mbox{DM}_{ij,k} & m_{i} \not= 0 \ \mbox{or} \ m_{k} \not= 0 
   & (5.16) \mbox{\ in \ \cite{Catani:2002hc}} \\ \hline
                      & 2. \ b & \mbox{D}_{ij}^{\ \ b} &  m_{i}=0 
   & (5.39) \mbox{\ in \ \cite{Catani:1996vz}} \\ \hline
                      &                    & \mbox{DM}_{ij}^{\ \ b} & m_{i} \not= 0 
   & (5.50) \mbox{\ in \ \cite{Catani:2002hc}} \\ \hline
   (2) \ (i,j)=(g,g)  & 3. \ k & \mbox{D}_{ij,k} &  m_{k}=0 
   & (5.9) \mbox{\ in \ \cite{Catani:1996vz}} \\ \hline
                      &                    & \mbox{DM}_{ij,k} & m_{k} \not= 0 
   & (5.19) \mbox{\ in \ \cite{Catani:2002hc}} \\ \hline
                      & 4. \ b & \mbox{D}_{ij}^{\ \ b} & --- 
   & (5.40) \mbox{\ in \ \cite{Catani:1996vz}} \\ \hline
   (3) \ (a,i)=(f,g)  & 5. \ k & \mbox{D}^{ai}_{\ \ k} & m_{k}=0 
   & (5.65) \mbox{\ in \ \cite{Catani:1996vz}} \\ \hline
                      &                    & \mbox{DM}^{ai}_{\ \ k} & m_{k} \not= 0 
   & (5.81) \mbox{\ in \ \cite{Catani:2002hc}} \\ \hline
                      & 6. \ b & \mbox{D}^{ai,b} & --- 
   & (5.145) \mbox{\ in \ \cite{Catani:1996vz}} \\ \hline
   (4) \ (a,i)=(g,g)  & 7. \ k & \mbox{D}^{ai}_{\ \ k} & m_{k}=0 
   & (5.68) \mbox{\ in \ \cite{Catani:1996vz}} \\ \hline
                      &                    & \mbox{DM}^{ai}_{\ \ k} & m_{k} \not= 0 
   & (5.85) \mbox{\ in \ \cite{Catani:2002hc}} \\ \hline
                      & 8. \ b & \mbox{D}^{ai,b} &  --- 
   & (5.148) \mbox{\ in \ \cite{Catani:1996vz}} \\ \hline
\mbox{Dipole 2}    &  &  & & 
\\ \hline
(5) \ (i,j)=(f,\bar{f})  &  & & &  
\\ \hline
\phantom{(5)} \ (i,j)=(u,\bar{u}) & 9u. \ k & \mbox{D}_{ij,k} & m_{k}=0 
   & (5.8) \mbox{\ in \ \cite{Catani:1996vz}} \\ \hline
\phantom{(5)} \                 &                    & \mbox{DM}_{ij,k} & m_{k} \not= 0 
   & (5.17) \mbox{\ in \ \cite{Catani:2002hc}} \\ \hline
                      & 10u. \ b & \mbox{D}_{ij}^{\ \ b} &  --- 
   & (5.41) \mbox{\ in \ \cite{Catani:1996vz}} \\ \hline
\phantom{(5)} \ (i,j)=(t,\bar{t}) & 9t. \ k & \mbox{DM}_{ij,k}   & --- 
   & (5.17) \mbox{\ in \ \cite{Catani:2002hc}} \\ \hline
\phantom{(5)} \  & 10t. \ b & \mbox{DM}_{ij}^{\ \ b} &  --- 
   & (5.51) \mbox{\ in \ \cite{Catani:2002hc}} \\ \hline
\mbox{Dipole 3}    &  &  & & 
\\ \hline
(6) \ (a,i)=(f,f)  &  & &  & 
\\ \hline
\phantom{(6)} \ (a,i)=(u,u) & 11u. \ k & \mbox{D}^{ai}_{\ \ k}   & m_{k}=0 
   & (5.67) \mbox{\ in \ \cite{Catani:1996vz}} \\  \hline
\phantom{(6)} \  &   & \mbox{DM}^{ai}_{\ \ k} & m_{k} \not= 0 
   & (5.83) \mbox{\ in \ \cite{Catani:2002hc}} \\  \hline
                      & 12u. \ b & \mbox{D}^{ai,b} &  --- 
   & (5.147) \mbox{\ in \ \cite{Catani:1996vz}} \\  \hline
\mbox{Dipole 4}  &  &  & & 
\\  \hline
(7) \ (a,i)=(g,f)  &  &  & & 
\\  \hline
\phantom{(7)} \ (a,i)=(g,u) & 13u. \ k & \mbox{D}^{ai}_{\ \ k}   & m_{k}=0 
   & (5.66) \mbox{\ in \ \cite{Catani:1996vz}} \\ \hline
\phantom{(7)} \   &       & \mbox{DM}^{ai}_{\ \ k} & m_{k} \not= 0 
   & (5.82) \mbox{\ in \ \cite{Catani:2002hc}} \\ \hline
                      & 14u. \ b & \mbox{D}^{ai,b} &  --- 
   & (5.146) \mbox{\ in \ \cite{Catani:1996vz}} \\ \hline 
\end{array} \nonumber
\end{eqnarray}
\caption{\small
\label{tab:creation-order-dip}
The emitter pair is denoted $(i,j)$ or $(a,i)$ and, respectively, the spectator $k(\not=i,j)$ or $b(\not=a)$. 
$\mbox{D}$ is a massless dipole and $\mbox{DM}$ is a massive dipole.
The equation numbers refer to \cite{Catani:1996vz} for massless dipoles ($\mbox{D}$) and
to \cite{Catani:2002hc} for massive ones ($\mbox{DM}$).
Dipole 2 has the other flavors, massless down and massive bottom quarks.
Dipole 3 and 4 have also the other light flavors, $f=\bar{u},d,\bar{d}$.
}
\end{table}

\begin{table}[h!]
  \centering
\begin{eqnarray}
\begin{array}{|l|l|l|l|l|} \hline
\ \ \ \ \mbox{Emitter}    & \mbox{Spectator} & \mbox{I-terms} & 
\ \  \mbox{Condition} & \ \mbox{Eq.} \\ \hline
   (1) \ i=f  & 1. \ k & \mbox{I}_{f}(i,k)   & m_{i}=m_{k}=0 
   & (\mbox{C}.27) \mbox{\ in \ \cite{Catani:1996vz}} \\ \hline
                      &                    & \mbox{IM1}(i,k) & m_{i} \not= 0 \ \mbox{or}  \ m_{k} \not= 0 
   & (6.16) \mbox{\ in \ \cite{Catani:2002hc}} \\ \hline
                      & 2. \ b & \mbox{I}_{f}(i,b) &  m_{i}=0 
   & (\mbox{C}.27) \mbox{\ in \ \cite{Catani:1996vz}} \\ \hline
                      &                    & \mbox{IM2}(i,b) & m_{i} \not= 0 
   & (6.52) \mbox{\ in \ \cite{Catani:2002hc}} \\ \hline
   (2) \ i=g  & 3. \ k & \mbox{I}_{g}(i,k) &  m_{k}=\{\mbox{m}_{F}\}=0 
   & (\mbox{C}.27) \mbox{\ in \ \cite{Catani:1996vz}} \\ \hline
                      &                    & \mbox{IM3}(i,k) & m_{k} \not= 0 \ \mbox{or} \ \{\mbox{m}_{F}\} \not=0 
   & (6.16) \mbox{\ in \ \cite{Catani:2002hc}} \\ \hline
                      & 4. \ b & \mbox{I}_{g}(i,b) & \{\mbox{m}_{F}\} =0 
   & (\mbox{C}.27) \mbox{\ in \ \cite{Catani:1996vz}} \\ \hline
                      &                    & \mbox{IM4}(i,b) & \{\mbox{m}_{F}\} \not=0 
   & (6.52) \mbox{\ in \ \cite{Catani:2002hc}} \\ \hline
   (3) \ a=f  & 5. \ k & \mbox{I}_{f}(a,k) & m_{k}=0 
   & (\mbox{C}.27) \mbox{\ in \ \cite{Catani:1996vz}} \\ \hline
                      &                    & \mbox{IM5}(a,k) & m_{k} \not= 0 
   & (6.52) \mbox{\ in \ \cite{Catani:2002hc}} \\ \hline
                      & 6. \ b & \mbox{I}_{f}(a,b) & --- 
   & (\mbox{C}.27) \mbox{\ in \ \cite{Catani:1996vz}} \\ \hline
   (4) \ a=g  & 7. \ k & \mbox{I}_{g}(a,k) & m_{k}=0 
   & (\mbox{C}.27) \mbox{\ in \ \cite{Catani:1996vz}} \\ \hline
                      &                    & \mbox{IM7}(a,k) & m_{k} \not= 0 
   & (6.52) \mbox{\ in \ \cite{Catani:2002hc}} \\ \hline
                      & 8. \ b & \mbox{I}_{g}(a,b) &  --- 
   & (\mbox{C}.27) \mbox{\ in \ \cite{Catani:1996vz}} \\ \hline
\end{array} \nonumber
\end{eqnarray}
\caption{\small
\label{tab:creation-order-I}
The emitter is denoted $i$ or $a$ and, respectively, the spectator $k(\not=i)$ or $b(\not=a)$. 
$\mbox{I}$ is the massless insertion operator and $\mbox{IM}$ is a massive one. 
The equation numbers refer to \cite{Catani:1996vz} for massless insertions ($\mbox{I}$) and
to \cite{Catani:2002hc} for massive ones ($\mbox{IM}$).
The set of the heavy-quark masses the running in the quark loop are abbreviated $\{\mbox{m}_{F}\}$.
}
\end{table}

\begin{table}[h!]
  \centering
\begin{eqnarray}
\begin{array}{|l|l|l|l|l|} \hline
\mbox{Emitter pair}    & \mbox{Spectator} & \mbox{P and K} & 
\ \  \mbox{Condition} & \ \mbox{Eq.} \\ \hline
\mbox{Dipole 1}    &  &  & & \\ \hline
   (3) \ (a,a',i)=(f,f,g)  &                   & \mbox{K}^{aa'}_{0} & --- 
   & (\mbox{C}.17) \mbox{\ in \ \cite{Catani:1996vz}} \\ \hline
                      & 5. \ k & \mbox{P}^{aa'}_{\ \ k} & --- 
   & (\mbox{C}.29) \mbox{\ in \ \cite{Catani:1996vz}} \\ \hline
                      &                    & \mbox{K}^{aa'}_{\ \ k} & m_{k} =
                      0 
   & (\mbox{C}.31) \mbox{\ in \ \cite{Catani:1996vz}} \\ \hline
                      &                    & \mbox{KM}^{aa'}_{\ \ k} & m_{k} \not= 0 \ \mbox{or} \ \{\mbox{m}_{F}\} \not=0 
   & (6.55) \mbox{\ in \ \cite{Catani:2002hc}} \\ \hline
                      & 6. \ b & \mbox{P}^{aa',b} & --- 
   & (\mbox{C}.29) \mbox{\ in \ \cite{Catani:1996vz}} \\ \hline
                      &                    & \mbox{K}^{aa',b} & --- 
   & (\mbox{C}.33) \mbox{\ in \ \cite{Catani:1996vz}} \\ \hline
   (4) \ (a,a',i)=(g,g,g)  &                    & \mbox{K}^{aa'}_{0} & --- 
   & (\mbox{C}.18) \mbox{\ in \ \cite{Catani:1996vz}} \\ \hline
                      & 7. \ k & \mbox{P}^{aa'}_{\ \ k} & --- 
   & (\mbox{C}.29) \mbox{\ in \ \cite{Catani:1996vz}} \\ \hline
                      &                    & \mbox{K}^{aa'}_{\ \ k} & m_{k} = 0 
   & (\mbox{C}.31) \mbox{\ in \ \cite{Catani:1996vz}} \\ \hline
                      &                    & \mbox{KM}^{aa'}_{\ \ k} & m_{k} \not= 0 \ \mbox{or} \ \{\mbox{m}_{F}\} \not=0 
   & (6.55) \mbox{\ in \ \cite{Catani:2002hc}} \\ \hline
                      & 8. \ b & \mbox{P}^{aa',b} & --- 
   & (\mbox{C}.29) \mbox{\ in \ \cite{Catani:1996vz}} \\ \hline
                      &                    & \mbox{K}^{aa',b} & --- 
   & (\mbox{C}.33) \mbox{\ in \ \cite{Catani:1996vz}} \\ \hline
\mbox{Dipole 3}    &  &  & & \\ \hline
(6) \ (a,a',i)=(f,g,f)  &  & &  & \\ \hline
\phantom{(6)} \ (a,a',i)=(u,g,u) &                     & \mbox{K}^{aa'}_{0} & --- 
   & (\mbox{C}.15) \mbox{\ in \ \cite{Catani:1996vz}} \\  \hline
                      & 11u. \ k & \mbox{P}^{aa'}_{\ \ k} & --- 
   & (\mbox{C}.29) \mbox{\ in \ \cite{Catani:1996vz}} \\  \hline
                      &                    & \mbox{KM}^{aa'}_{\ \ k} & m_{k} \not= 0 
   & (6.55) \mbox{\ in \ \cite{Catani:2002hc}} \\ \hline
                      & 12u. \ b & \mbox{P}^{aa',b} & --- 
   & (\mbox{C}.29) \mbox{\ in \ \cite{Catani:1996vz}} \\  \hline
                      &                    & \mbox{K}^{aa',b} & --- 
   & (\mbox{C}.33) \mbox{\ in \ \cite{Catani:1996vz}} \\ \hline
\mbox{Dipole 4}  &  &  & & \\  \hline
(7) \ (a,a',i)=(g,f,\bar{f})  &  &  & & \\  \hline
   \phantom{(6)} \ (a,a',i)=(g,u,\bar{u}) &                     & \mbox{K}^{aa'}_{0} & --- 
   & (\mbox{C}.16) \mbox{\ in \ \cite{Catani:1996vz}} \\ \hline
                      & 13u. \ k & \mbox{P}^{aa'}_{\ \ k} & --- 
   & (\mbox{C}.29) \mbox{\ in \ \cite{Catani:1996vz}} \\ \hline
                      &                    & \mbox{KM}^{aa'}_{\ \ k} & m_{k} \not= 0 
   & (6.55) \mbox{\ in \ \cite{Catani:2002hc}} \\ \hline
                      & 14u. \ b & \mbox{P}^{aa',b} & --- 
   & (\mbox{C}.29) \mbox{\ in \ \cite{Catani:1996vz}} \\ \hline
                      &                    & \mbox{K}^{aa',b} & --- 
   & (\mbox{C}.33) \mbox{\ in \ \cite{Catani:1996vz}} \\ \hline
\end{array} \nonumber
\end{eqnarray}
\caption{\small
\label{tab:creation-order-pandk}
The emitter is denoted $a'$ and different from spectator $b(\not=a')$. 
The massless case is abbreviated $\mbox{K}$ and massive one $\mbox{KM}$. 
The equation numbers refer to \cite{Catani:1996vz} for massless insertions ($\mbox{K}$) and
to \cite{Catani:2002hc} for massive ones ($\mbox{KM}$).
For the cases Dipole 3 and 4, $f$ can take the light flavors $f=u,\bar{u},d,\bar{d}$ and so on.
}
\end{table}

\vspace*{20mm}
\newpage
\vspace*{20mm}
\newpage

\section{Correlation functions of gluon emitter}
\label{sec:appB}
\renewcommand{\theequation}{\ref{sec:appB}.\arabic{equation}}
\setcounter{equation}{0}

We use the following definitions of spinor products~\cite{Xu:1986xb}:
\begin{eqnarray}
\bra k_{1} k_{2} \ket &=& \sqrt{k_{1}^{-} k_{2}^{+}} e^{+i \phi(k_{1})} - \sqrt{k_{1}^{+} k_{2}^{-}} 
e^{+i \phi(k_{2})} \\
k^{\pm} &=& |\vec{k}| \pm k_{z} \\
e^{+i \phi(k)} &=& \frac{k_{\bot}}{|k_{\bot}|} \\
k_{\bot} &=& k_{x} + i k_{y} 
\end{eqnarray}
If the reduced momentum of the initial gluon emitter 
lies in z-axis with $l=E_{l}(1,0,0,\pm1)$ 
the spinor product are defined as $\bra p_{i} \tilde{p}_{ai} \ket_{z}$ 
to avoid an ambiguity in the intermediate steps of the calculation,
\begin{eqnarray}
\label{eq:braketz}
\bra k l \ket_{z} &=& \left\{ \begin{array}{cc}
\sqrt{2 E_{l} k^{-}}e^{+i \phi(k)} & (E_{l} = l_{z})  \\
-\sqrt{2 E_{l} k^{+}} & (E_{l} = -l_{z})
\end{array}\right.  \
 \\
{}_z\!\!\bra l_{1} l_{2} \ket_{z} &=& \left\{ \begin{array}{cc}
-2\sqrt{E_{1} E_{2}}& (E_{1} = l_{1z} \ \mbox{and} \ E_{2} = -l_{2z})  \\
2\sqrt{E_{1} E_{2}} & (E_{1} = -l_{1z} \ \mbox{and} \ E_{2} = l_{2z})
\end{array}\right.
\end{eqnarray}

\vspace{5mm}

The explicit forms of $e^{+i\phi(k)}\ \mbox{E}_{+}$ in \Eq{eq:dipcir1s} are shown for all massless and massive
dipoles with a gluon emitter. Here we write the quantity including the phase difference as $\mbox{E}_{+}'=e^{+i\phi(k)}\ \mbox{E}_{+}$.
The reduced momenta are ones used in the corresponding equations in \cite{Catani:1996vz,Catani:2002hc}, which are shown 
in Table~\ref{tab:creation-order-dip}. The $\tilde{p}_{ij1}$ and $\tilde{p}_{ai}$ are massless and 
the $\tilde{pM}_{ij}$ are massive ones.

\noindent
\subsection*{Massless momenta}

[Dipole 1-(2)-3($m_{k}=0$) and Dipole 2-(5)-9u($m_{k}=0$)]
\begin{eqnarray}
E1_{+}' = e^{+i \phi(\tilde{p}_{ij})}\frac{z_{i}\bra p_{j} p_{i} \ket\bra \tilde{p}_{ij} p_{i} \ket^{\star}}
{\sqrt{2} \bra p_{j} \tilde{p}_{ij} \ket}  
\label{E1}
\end{eqnarray}

[Dipole 1-(2)-4 and Dipole 2-(5)-10u($m_{k}=0$)]
\begin{eqnarray}
E2_{+}' = e^{+i \phi(\tilde{p}_{ij})}\frac{z_{i}\bra p_{j} p_{i} \ket\bra \tilde{p}_{ij} p_{i} \ket^{\star}}
{\sqrt{2} \bra p_{j} \tilde{p}_{ij} \ket} 
\end{eqnarray}

[Dipole 1-(4)-7($m_{k}=0$) and Dipole 3-(6)-11u($m_{k}=0$)]
\begin{eqnarray}
E3_{+}' = \frac{\bra p_{i} p_{k} \ket\bra p_{k}  \tilde{p}_{ai} \ket_{z}^{\star}}
{\sqrt{2}(1-u_{i}) \bra p_{i} \tilde{p}_{ai} \ket_{z}} 
\end{eqnarray}

[Dipole 1-(4)-8 and Dipole 3-(6)-12u]
\begin{eqnarray}
E4_{+}' = \frac{-s_{ai} \bra p_{i} p_{b} \ket_{z} {}_z\!\!\bra \tilde{p}_{ai}p_{b} \ket_{z}}
{\sqrt{2}s_{ab} \bra p_{i} \tilde{p}_{ai} \ket_{z}} 
\end{eqnarray}

\noindent
\subsection*{Massive momenta}

[Dipole 1-(2)-3($(m_{k}\not=0)$) and Dipole 2-(5)-9u($(m_{k}\not=0)$)]
\begin{eqnarray}
EM1_{+}' = e^{+i \phi(\tilde{pM}_{ij})}\frac{z_{i}^{(m)} \bra p_{j} p_{i} \ket
\bra \tilde{pM}_{ij} p_{i} \ket^{\star}}{\sqrt{2} \bra p_{j} \tilde{pM}_{ij} \ket}
\end{eqnarray}

\hspace{10mm}[Dipole 2-(5)-9t]
\begin{eqnarray}
EM1_{9t +}' = e^{+i \phi(\tilde{pM}_{ij})}\frac{z_{i}^{(m)} \bra p_{j}^{\flat} p_{i}^{\flat} \ket
\bra \tilde{pM}_{ij} p_{i}^{\flat} \ket^{\star}}{\sqrt{2} \bra p_{j}^{\flat} \tilde{pM}_{ij} \ket}
\end{eqnarray}
\hspace{30mm} with flat momentum
\begin{eqnarray}
p_{i(j)}^{\flat} = p_{i(j)} - \frac{m_{i(j)}^{2}}{2p_{i(j)} \cdot \tilde{pM}_{ij}} \tilde{pM}_{ij}
\end{eqnarray}

[Dipole 2-(5)-10t]
\begin{eqnarray}
EM2_{+}' = e^{+i \phi(\tilde{pM}_{ij})}\frac{z_{i} \bra p_{j}^{\flat} p_{i}^{\flat} \ket
\bra \tilde{pM}_{ij} p_{i}^{\flat} \ket^{\star}}{\sqrt{2} \bra p_{j}^{\flat} \tilde{pM}_{ij} \ket}  \label{EM2}
\end{eqnarray}
\hspace{30mm} with
\begin{eqnarray}
p_{i(j)}^{\flat} = p_{i(j)} - \frac{m_{i(j)}^{2}}{2p_{i(j)} \cdot \tilde{pM}_{ij}} \tilde{pM}_{ij}
\end{eqnarray}

[Dipole 1-(4)-7($(m_{k}\not=0)$) and Dipole 3-(6)-11u($(m_{k}\not=0)$)]
\begin{eqnarray}
EM3_{+}' = \frac{\bra p_{i} p_{k}^{\flat} \ket\bra p_{k}^{\flat}  \tilde{p}_{ai} \ket_{z}^{\star}}
{\sqrt{2}(1-u_{i}) \bra p_{i} \tilde{p}_{ai} \ket_{z}} 
\end{eqnarray}
\hspace{30mm} with
\begin{eqnarray}
p_{k}^{\flat} = p_{k} - \frac{m_{k}^{2}}{2p_{k} \cdot \tilde{p}_{ai}} \tilde{p}_{ai}
\end{eqnarray}

\newpage

\section{Comparison with the $t \bar{t}$ + 1jet}
\label{sec:appC}
\renewcommand{\theequation}{\ref{sec:appC}.\arabic{equation}}
\setcounter{equation}{0}

\begin{table}[h!]
  \begin{center}
    \leavevmode
    \begin{tabular}[h]{|c|l|l|}
      \hline

      &$b_0 [\GeV^{-4}]$ &$d_0 [\GeV^{-4}]$ \\
      \hline\multicolumn{3}{|c|}
      {$ \Pg(p_a)\Pg(p_b) \to \Pt(p_{\Pt}) \bar \Pt(p_{\bar\Pt}) \Pg(p_c) \Pg(p_d)$}  \\ \hline
      \Vone  & $ 7.82039670869613 \cdot 10^{-10}$ & $ 1.02594003852407\cdot 10^{-9}$ \\
      \Vtwo  & $ 7.82039670869604(1) \cdot 10^{-10}$ & $ 1.02594003852396(2)\cdot 10^{-9}$ \\
      \hline\multicolumn{3}{|c|}
      {$ q(p_a)\bar q(p_b) \to \Pt(p_{\Pt}) \bar \Pt(p_{\bar\Pt}) \Pg(p_c) \Pg(p_d) $} \\ \hline
      \Vone  & $ 1.12077211361620 \cdot 10^{-10}$ & $ 1.22619016939900 \cdot 10^{-10}$ \\
      \Vtwo  & $ 1.12077211361619(0) \cdot 10^{-10}$ & $ 1.22619016939909(1) \cdot 10^{-10}$ \\
      \hline\multicolumn{3}{|c|}
      {$ q(p_a)\Pg(p_b) \to \Pt(p_{\Pt}) \bar \Pt(p_{\bar\Pt}) \Pg(p_c) q(p_d) $} \\ \hline
      \Vone  & $ 2.75641273146785 \cdot 10^{-11}$ & $ 4.79768338384667 \cdot 10^{-11}$ \\
      \Vtwo  & $ 2.75641273146783(0) \cdot 10^{-11}$ & $ 4.79768338384625(3) \cdot 10^{-11}$ \\
      \hline\multicolumn{3}{|c|}
      {$ \bar q(p_a)\Pg(p_b) \to \Pt(p_{\Pt}) \bar \Pt(p_{\bar\Pt}) \bar q(p_c) \Pg(p_d)$} \\ \hline
      \Vone  & $ 3.46150168295956 \cdot 10^{-11}$ & $ 8.34555795894942 \cdot 10^{-11}$ \\
      \Vtwo  & $ 3.46150168295954(1) \cdot 10^{-11}$ & $ 8.34555795894963(2) \cdot 10^{-11}$ \\
      \hline\multicolumn{3}{|c|}
      {$ \Pg(p_a)\Pg(p_b) \to \Pt(p_{\Pt}) \bar \Pt(p_{\bar\Pt}) \bar q(p_c) q(p_d) $} \\ \hline
      \Vone  & $ 1.21420520114780 \cdot 10^{-11}$ & $ 2.13553289076589 \cdot 10^{-11}$ \\
      \Vtwo  & $ 1.21420520114779(0) \cdot 10^{-11}$ & $ 2.13553289076550(3) \cdot 10^{-11}$ \\
      \hline\multicolumn{3}{|c|}
      {$ q(p_a)\bar q(p_b) \to \Pt(p_{\Pt}) \bar \Pt(p_{\bar\Pt}) \bar q(p_c) q(p_d) $} \\ \hline
      \Vone  & $ 5.13710959990068 \cdot 10^{-12}$ & $ 9.06330902408356 \cdot 10^{-12}$ \\
      \Vtwo  & $ 5.13710959990064(1) \cdot 10^{-12}$ & $ 9.06330902408275(3) \cdot 10^{-12}$ \\
      \hline\multicolumn{3}{|c|}
      {$ u(p_a) \bar u(p_b) \to W^{+}(p_{w^{+}}) W^{-}(p_{w^{-}}) g(p_c) g(p_d) $}  \\ \hline
      \Vone  & $ 0.627402537098012 \cdot 10^{-9}$ & $  0.114149934878320\cdot 10^{-8}$ \\
      Ref.~\cite{Dittmaier:2007th} & $ 0.627402537098007 \cdot 10^{-9}$ & $  0.114149934878319 \cdot 10^{-8}$ \\
      \hline

   \end{tabular}
    \caption{
      The results of the $|\mbox{M}|^{2}$ and the sum of all dipole terms are shown.
      The two implementations agree at least to 14 digits for the matrix elements squared and at least to 12 digits
      for the sum of the subtraction terms.
      The phase space point used in the last entry corresponds to the first one in 
      {\tt ./lib/check/check5\_uux\_w+w-gg/inputm\_uux\_w+w-gg.h}.
      }
      \label{tab:real1}
  \end{center}
\end{table}

\begin{sidewaystable}[htbp]
  \begin{center}
    \leavevmode
    \begin{tabular}[h]{|c|l|l|l|}
      \hline
      & \ \ \ $c_{-2}$ & \ \ \ $c_{-1}$ & \ \ \ $c_{0}$ \\
      \hline
      \multicolumn{4}{|c|}
      {$gg  \to t\bar{t}g$}\\
      \hline
      
     \Vone & $2.49467966948003 \cdot 10^{-4}$ & $3.68989776683705 \cdot 10^{-4}$ & $-4.05387364353899 \cdot 10^{-4}$ \\
     \Vtwo & $2.49467966948004(1) \cdot 10^{-4}$ & $3.68989776683706(1) \cdot 10^{-4}$ & $-4.05387364353900(1) \cdot 10^{-4}$ \\
      \hline
      \multicolumn{4}{|c|}
     {$u\bar{u}  \to t\bar{t}g$} \\
      \hline
     \Vone & $1.38499897972387 \cdot 10^{-5}$ & $2.88738914389178 \cdot 10^{-5}$ & $-1.56576469322102 \cdot 10^{-5}$ \\
     \Vtwo & $1.38499897972387(0) \cdot 10^{-5}$ & $2.88738914389179(1) \cdot 10^{-5}$ & $-1.56576469322102(0) \cdot 10^{-5}$ \\
      \hline
      \multicolumn{4}{|c|}
     {$ug  \to t\bar{t}u$}\\
      \hline
     \Vone & $3.84580760674706 \cdot 10^{-6}$ & $7.73777480040817 \cdot 10^{-6}$ & $-5.19929995897616 \cdot 10^{-6}$ \\
     \Vtwo & $3.84580760674706(0) \cdot 10^{-6}$ & $7.73777480040817(1) \cdot 10^{-6}$ & $-5.19929995897616(1) \cdot 10^{-6}$ \\
      \hline
      \multicolumn{4}{|c|}
     {$g\bar{u}  \to t\bar{t}\bar{u}$}\\
      \hline
     \Vone & $6.22738241305372 \cdot 10^{-5}$ & $6.81530745255038 \cdot 10^{-5}$ & $-1.52377227863896 \cdot 10^{-4}$  \\
     \Vtwo & $6.22738241305372(0) \cdot 10^{-5}$ & $6.81530745255037(1) \cdot 10^{-5}$ & $-1.52377227863896(0) \cdot 10^{-4}$  \\
      \hline
    \end{tabular}
    \caption{
      The coefficients for the color and spin averaged results for the {\bf I}-operator. 
      At least 14 digits agreements are obtained.}
    \label{tab:virtual1}
  \end{center}
\end{sidewaystable}

\end{document}